\documentclass[a4paper, 12pt, oneside]{article}
\usepackage[cp1251]{inputenc}
\usepackage[english, russian]{babel}
\usepackage{ graphics,amsfonts, amsmath,bm,amssymb, array,eucal}
\usepackage[dvips]{graphicx}
\usepackage[flushleft,small]{caption2}
\renewcommand{\Re}{\mathop{\rm Re\,}}
\renewcommand{\Im}{\mathop{\rm Im\,}}

\makeatother {
\newcommand{\mc}[1]{\mathcal{#1}}
\newcommand{\E}{\mc{E}}

\begin{document}
\thispagestyle{empty} \large
\renewcommand{\abstractname}{\, }

 \begin{center}
\bf The Kramers Problem for Quantum Fermi Gases with Velocity--Dependent
Collision Frequency and Diffusive Boundary Conditions
\end{center}\medskip
\begin{center}
  \bf A. Yu. Kvashnin\footnote{$malin_85@mail.ru$},
  A. V. Latyshev\footnote{$avlatyshev@mail.ru$} and
  A. A. Yushkanov\footnote{$yushkanov@inbox.ru$}
\end{center}\medskip

\begin{center}
{\it Faculty of Physics and Mathematics,\\ Moscow State Regional
University, 105005,\\ Moscow, Radio str., 10--A}
\end{center}\medskip

\begin{abstract}
    The classical Kramers problem of the kinetic theory is analytically solved.
The Kramers problem about isothermal sliding for quantum Fermi
gases is considered. Quantum gases with the velocity--dependent
collision frequency are considered. Diffusive  boundary conditions are applied.
Dependence of isothermal sliding  on the resulted chemical potential 
    is found out.

{\bf Key words:} statement of problem, dispersion function,
eigenvalues, eigen\-functions, expansion by eigenfunctions,
collisional rarefied gas, boundary value Riemann problem,
singular integral equation, exact solution.

PACS numbers: 05.20.Dd Kinetic theory, 47.45.-n Rarefied gas dynamics,
02.30.Rz Integral equations.
\end{abstract}

\tableofcontents
\setcounter{secnumdepth}{4}

\begin{center}
\section{Введение}
\end{center}
\markboth{}{ВВЕДЕНИЕ}
Динамика разреженного газа, большинство ранних исследований которой в
начале прошлого века касались, в основном, либо течений с очень
малой скоростью \cite{91}, либо различного рода "внутренних"\,
течений (в трубах, соплах, насадках и т.д.), связанных с
проблемами получения глубокого вакуума, претерпела в середине
прошлого столетия свое второе рождение, что обусловлено было в
первую очередь развитием сверхзвуковой высотной авиацией,
созданием ракетно--космической техники, разработкой новых
химических технологий. Этим объясняется и то пристальное
внимание, которое привлекает к себе в настоящее время
классическая кинетическая теория, созданная в XIX веке
выдающимся австрийским физиком Людвигом Больцманом
\cite{4} и \cite{4a}.

    Но исторической датой возникновения кинетической теории газов
следует считать 1859 год, когда Максвелл на заседании Британской
ассоциации содействия развитию науки прочитал свой доклад, в
котором был впервые использован статистический подход к
проблеме. В 1860 году в серии из двух работ \cite{128} Максвелл
опубликовал результаты исследований, в которых установил закон
распределения скоростей молекул в смеси газов (так называемое
максвелловское распределение по скоростям) и закон
равнораспределения средней энергии молекул в смеси газов. Эти
результаты были впоследствии (в 1867 г.) уточнены и улучшены
Максвеллом в работе \cite{127}, посвященной кинетической теории
неоднородных газов. В ней Максвелл вывел уравнения переноса,
определяющие полную скорость изменения любой средней величины,
характеризующей то или иное молекулярное свойство.

     В 1875 году Кундт и Варбург
экспериментально доказали, что в достаточно
  разреженном газе
 молекулы "скользят"\, вдоль стенок. Скорость скольжения
  пропорциональна градиенту массовой скорости
вне слоя Кнудсена. Коэффициент пропорциональности при этом
называется коэффициентом изотермического скольжения. Кундт и
Варбург нашли, что коэффициент изотермического скольжения
оказался порядка длины свободного пробега молекул и обратно
пропорционален давлению. Анализ, проведенный Максвеллом,
основывался на предположении, что распределение падающих молекул
вблизи стенки не отличается от распределения в прилегающем
объеме газа. В действительности же молекулы газа перед тем, как
удариться о поверхность, испытывают в среднем не менее одного
соударения с молекулами, покидающими поверхность и имеющими (при
полной аккомодации) нормальное максвелловское распределение
скоростей, в отличие от того модифицированного распределения
скоростей, которое характерно для объема газа при наличии
градиента температуры.

    Кинетическое уравнение, выведенное в 1872
году \cite{4a} австрийским физиком Людвигом Больцманом и носящее
его имя, обладает громадным физическим содержанием. Для решения
кинетического уравнения Больцмана в первой половине прошлого
века было предложено несколько различных методов и большое число
их различных вариаций.

    В 1912 году Гильберт предложил метод
последовательных приближений (обычный метод малого параметра),
который позволил свести решение кинетического уравнения к
решению рекуррентной системы линейных неоднородных интегральных
уравнений. Он показал, что уравнение Больцмана эквивалентно
интегральному уравнению Фредгольма второго рода, для которого
оказалось возможным построить строгую математическую теорию.
Таким образом, Гильберт смог доказать существование и
единственность решения и установить некоторые из его свойств.

    Затем в 1917 году Энског в своей докторской диссертации
\cite{118} независимо от него Чепмен в работе \cite {110a}
предложили другой метод решения уравнения Больцмана, который
является некоторой модификацией метода Гильберта и который
позволяет вывести как уравнение Эйлера, так и уравнение
Навье-Стокса. В методе Чепмена --- Энского \cite{92}, малый
параметр по которому ведется разложение, входит в решение более
сложным, вообще говоря, не аналитическим образом, поэтому при
одинаковом числе членов разложения, входит в решение, полученное
по методу Чепмена --- Энского \cite{92} может оказаться более
точным, чем то, что было получено с использованием метода
Гильберта. Хотя имеются примеры и того \cite{92}, что уравнения,
полученные по методу Чепмена --- Энского, могут не иметь
решения, в то время как метод Гильберта позволяет построить
решение в любом приближении. Следует отметить \cite{91}, что
метод, предложенный Чепменом и Энскогом, не только позволил
обоснованно вывести уравнение Эйлера и Навье --- Стокса и их
аналогов для релаксирующих сред, но и дал возможность установить
область их применимости, снабдив их правильными начальными и
граничными условиями и коэффициентами переноса.

    Основная трудность, возникающая при решении задач
аэродинамики разреженных газов с использованием кинетического
уравнения Больцмана, заключается в сложности самого этого
уравнения, в особенности интеграла столкновений, стоящего в его
правой части.

    Поэтому \cite{93}, начиная с 60-х годов XX века стал
развиваться новый подход к исследованию уравнения Больцмана,
суть которого состоит в следующем: используя определенные
закономерности столкновений молекул, проводят упрощение правой,
интегральной части уравнения Больцмана, а полученное таким
образом уравнение рассматривается как его "модель" \,, которая,
сохраняя основные качественные особенности исходного уравнения,
является в тоже время значительно более простым выражением.

    Первая из статистических моделей уравнения Больцмана была
предложена независимо друг от друга в работах \cite{103},
\cite{144}. Соображения, которые привели к этой модели весьма
просты \cite{96}.

    А именно, поскольку больцмановский оператор столкновений
представляет собой скорость стремления функции распределения к
равновесной максвелловской, то можно попытаться в общих чертах
представить эту скорость в виде отношения разности
действительной функции и равновесной к некоторому характерному
времени затухания начальных возмущений в однородном газе. В
результате получается следующее модельное кинетическое
уравнение:
 $$
\dfrac {\partial f}{\partial t}+ (\mathbf{v} \nabla)f=
\dfrac{f_{eq}-f}{\tau},
$$
где $f_{eq}$ - локально равновесная максвелловская функция
распределения. В литературе его называют моделью БГК (Бхатнагар,
Гросс, Крук), БГК -- уравнением  или
просто релаксационным кинетическим уравнением.

    История точных решений модельных кинетических уравнений
начинается с 1960 года, когда Кейз в работе \cite{104} ввел в
рассмотрение метод решения уравнений переноса, состоящий в
разложении решения по дискретным и сингулярным обобщенным
собственным функциям соответствующего оператора переноса и
нахождении коэффициентов этого разложения с помощью техники
сингулярных интегральных уравнений. Этот метод стал источником
точных решений, получаемых аналитически в замкнутой форме.

 В
1962 году Черчиньяни в работе \cite{105} разработал метод Кейза
применительно к задачам кинетической теории, в частности, к
задаче о сдвиговом течении разреженного газа при постоянной
температуре (задача Крамерса). При этом, на основе
аналитического решения уравнения Больцмана с оператором
столкновений в форме БГК модели, была получена точная формула
для вычисления коэффициента изотермического скольжения газа
вдоль плоской твердой поверхности.

В работе \cite{105}
Черчиньяни решил задачу Крамерса с учетом аккомодации молекул, в
\cite{106a} рассмотрел нестационарный случай и в \cite{106}
выяснил зависимость коэффициента скольжения от частоты
столкновений.

В работе \cite{30} был предложен принципиально новый
математический подход, который позволяет получить точные решения
линеаризованных уравнений Больцмана с оператором столкновений
БГК-- или эллипсоидально--статистической модели путем сведения
их к интегро--дифференциальным уравнениям типа свертки.
Полученные таким образом уравнения преобразованием Фурье
сводятся к краевым задачам Римана --- Гильберта и решаются затем
методами теории функций комплексного переменного
(\cite{8},\cite{9}, \cite{88}).

В работе А.В.  Латышева и А.А. Юшканова \cite{120a} впервые был
применен метод разложения решения по сингулярным собственным
функциям для решения системы двух интегро--дифференциальных
уравнений переноса. Значительное число аналитических решений
граничных задач для различных модельных кинетических уравнений
было получено в работах (\cite{31}-\cite{83}).
 Квантовые ферми--газы изучались,
главным образом, в рамках рассмотрения кинетики электронов в
полупроводниках и металлах (см., например, \cite{14g}).
Квантовые бозе--газы рассматривались при
исследовании кинетики фононов, магнонов и экситонов в
конденсированных средах \cite{14g}. Однако и электроны в твердых
телах, и фононы относятся к квазичастицам, чье поведение
отличается от поведения свободных частиц, а электроны, кроме
того, обладают зарядом, так что в этом случае мы имеем дело с
кинетикой плазмы.

В настоящей работе на основе нелинейного релаксационного кинетического уравнения
для квантовых ферми--газов с переменной частотой столкновений выводится
с использованием закона сохранения импульса линеаризованное уравнение 
для задач скольжения газа вдоль плоской твердой поверхности. Проводится 
постановка полупространственной задачи Крамерса о нахождении скорости
изотермического скольжения газа вдоль плоской поверхности. Затем развивается
математический аппарат, необходимый для решения задачи. Аналитическое решение
задачи строится с использованием техники сингулярных интегральных уравнений с 
ядром Коши и краевых задач теории функций комплексного переменного.
В конце работы строится профиль массовой скорости газа в полупространстве.

\begin{center}
\item{}\section{Нелинейное уравнение для квантовых ферми--газов}
\end{center}

\markboth{}{НЕЛИНЕЙНОЕ КИНЕТИЧЕСКОЕ УРАВНЕНИЕ ДЛЯ ФЕРМИ--ГАЗОВ}

Приведем вывод кинетического уравнения для задач скольжения
квантовых ферми--газов с переменной частотой столкновений.

Рассмотрим $\tau$--модельное уравнение Больцмана для случая,
когда частота столкновений молекул газа пропорциональна модулю
скорости молекул:
$$
\dfrac {\partial f}{\partial t}+ (\mathbf{v} \nabla)f= \nu
(\mathbf{v})(f_{M}^{*}-f).
\eqno{(2.1)}
$$

Здесь $f=f(\mathbf{r},\mathbf{v}, t)$ -- функция распределения газовых
молекул, $\nu(\mathbf{v})$ -- частота столкновений, зависящая от
модуля относительной скорости молекул,
$$
\nu(\mathbf {v})=\dfrac{|\mathbf {v}-\mathbf {u}|}{l},
$$
$\mathbf{u}=\mathbf {u}(\mathbf {r}, t)$ -- массовая скорость
газа, определяемая соотношением
$$
\mathbf {u}= \dfrac{1}{n(\mathbf{r},t)} \int \mathbf{v}
f(\mathbf{r},\mathbf{v}, t)d\Omega_{M},\qquad d\Omega_{M}=d^3v,
\eqno{(2.2)}
$$
$$
n(\mathbf{r},t)=\int f(\mathbf{r},\mathbf{v}, t)d\Omega_{M},
$$
$n(\mathbf{r},t)$ -- числовая плотность (концентрация) молекул
газа,
$ |\mathbf {v}-\mathbf {u}(\mathbf{r},t)|$ --
модуль скорости
молекулы в системе отсчета,
относительно которой газ в данной точке $\mathbf{r}$
покоится, т. е. имеет массовую
скорость, равную нулю, $\mathbf{v}$ -- молекулярная
скорость газа в лабораторной
системе отсчета, $\mathbf{u}(\mathbf{r},t)$ -- массовая
скорость газа в точке $\mathbf{r}$ в
лабораторной системе отсчета, $l$ -- средняя длина свободного
пробега молекул, $l=\tau v_T$, $\tau$ -- время релаксации,
т. е. время между двумя последовательными столкновениями молекул
газа, $v_T=1/\sqrt{\beta}$ -- тепловая скорость газа,
$\beta=\sqrt{m/2kT}$.

Функция $f_M^*$ -- аналог
максвелловской функции распределения, который
имеет следующий вид
$$
f_{M}^{*}(\mathbf {r},t)=n \Big(\dfrac{\beta}{\pi}\Big)^{3/2}
\exp \Big[-\beta(\mathbf {v}-\mathbf u_*(\mathbf r,t))^2\Big],
\eqno{(2.3)}
$$
где $ \mathbf{u}_*(\mathbf r,t)$ -- параметр модели, определяемый
из требования выполнения закона сохранения импульса:
$$
\int \nu (\mathbf {v})\mathbf {v}f(\mathbf{r},\mathbf{v},
t)d\Omega_{M}=\int \nu (\mathbf {v})\mathbf
{v}f_{M}^{*}d\Omega_{M}.
\eqno{(2.4)}
$$

Отметим, что локально равновесная максвелловская функция
распределения имеет следующий вид:
$$
f_{M}(\mathbf {r},t)=n \Big(\dfrac{\beta}{\pi}\Big)^{3/2}
\exp \Big[-\beta(\mathbf {v}-\mathbf u(\mathbf
r,t))^2\Big].
$$
Массовая скорость газа $u(\mathbf r,t)$ из $f_M$ вычисляется
согласно (2.2).

Рассмотрим обобщение уравнения (2.1) на случай квантовых
ферми--газов. Это обобщение состоит в том,
что аналог максвелловской
функции распределения $f_{M}^{*}$ заменяетcя на аналог функции
распределения Ферми --- Дирака $f_{F}^{*}$.
При этом для ферми--газов получаем уравнение вида (2.1):
$$
\dfrac {\partial f}{\partial t}+ (\mathbf{v} \nabla)f= \nu
(\mathbf{v})(f_{F}^{*}-f),
\eqno{(2.5)}
$$
в котором
$$
f_{F}^{*}=f_{F}^{*}(\mathbf{r},\mathbf{v},
t)=\dfrac{1}{1+\exp\Big[\beta(\mathbf {v}-\mathbf u_*(\mathbf r,t))^2-
{\mu}/{kT}\Big]}.
\eqno{(2.6)}
$$

Здесь в уравнении (2.6) величина $\mu$
(химический потенциал молекул) и температура $T$ считаются постоянными.
В уравнении (2.6) параметр модели $\mathbf{u}_*(\mathbf r,t)$
определяется законом сохранения импульса:
$$
\int \nu (\mathbf {v})\mathbf {v}f(\mathbf{r},\mathbf{v},
t)d\Omega=\int \nu (\mathbf {v})\mathbf
{v}f_{F}^{*}d\Omega.
\eqno{(2.7)}
$$
Здесь
$$
d\Omega=\dfrac{2s+1}{(2\pi \hbar)^3}d^3p,
%\eqno{(2.9)}
$$
$s$ -- спин частицы, $\mathbf {p}=m\mathbf {v}$ -- её
импульс, $\hbar$ -- постоянная Планка.

Кроме того, в выражении (2.6) химический потенциал
$\mu$ изменяется в пределах от $-\infty$ до $+\infty$ как
числовой параметр.

Будем считать, что массовая скорость газа $\mathbf{u}$ мала по
сравнению со звуковой:
$$
\dfrac{|\mathbf{u(\mathbf{r},t)|}}{v_T}\ll 1.
$$
\begin{center}
\item{}\section{Линеаризованное уравнение для квантовых ферми--газов}
\end{center}
\markboth{}{ЛИНЕАРИЗОВАННОЕ КИНЕТИЧЕСКОЕ УРАВНЕНИЕ}
Сначала проведем линеаризацию кинетического уравнения (2.5) для
квантовых ферми--газов.
Линеаризацию аналога функции распределения Ферми ---
Дирака проведем по параметру $\mathbf{u}_*$.

Ввёдем обозначение:$\E_{*}=\beta(\mathbf{v}-\mathbf{u_*})^2-{\mu}/{kT}$.
Линеаризация $f_{F}^{*}$ по величине $\mathbf {u_*}$
 приводит к следующему выражению:
$$
f_{F}^{*}=f_{F}^{*}\Big|_{\mathbf {u_*}=0}+\dfrac{\partial
f_{F}^{*}}{\partial \mathbf {u_*}}\Big|_{\mathbf {u_*}=0}
\mathbf {u_*},
%\eqno{(12)}
$$
или в явном виде,
$$
f_{F}^{*}=f_{F}(v)+2g(v)\beta\mathbf{v}\mathbf {u_*}.
\eqno{(3.1)}
$$

Здесь $f_{F}(v)$ -- абсолютное распределение Ферми ---
Дирака,
$$
f_{F} (v)=\dfrac{1}{1+\exp
(\beta v^2-{\mu}/{kT})},\;
%$$
%$$
g(v)=\dfrac{\exp(\beta v^2-{\mu}/{kT})}{\Big(1+
\exp(\beta v^2-{\mu}/{kT})\Big)^2}.
$$

Связь между двумя последними функциями выражается равенством:
$
g(v)=-\dfrac{\partial f_{F}(\E)}{\partial \E},
$
где $\E=\beta v^2-{\mu}/{kT}$.

Равенство (3.1) означает, что функцию распределения следует
искать в следующем линеаризованном относительно абсолютного
распределения Ферми --- Дирака виде:
$$
f(\mathbf{r},\mathbf{v},
t)=f_{F}(v+g(v)\varphi(\mathbf{r},\mathbf{v}, t).
 \eqno{(3.2)}
$$

Найдем разность выражений (3.1) и (3.2). Имеем:
$$
f^{*}_{F}-f=g(v)\Big[2\beta\mathbf{v}\mathbf{u_*}-
\varphi(\mathbf{r}, \mathbf v, t)\Big].
\eqno{(3.3)}
$$

В линейном приближении зависимость частоты столкновений молекул
от модуля скорости молекул линейная:
$$
\nu(\mathbf v)=\dfrac{v}{l},\qquad v=\sqrt{v_x^2+v_y^2+v_z^2}.
$$

Подставляя разность (3.3) в кинетическое уравнение (2.5),
получаем следующее уравнение:
$$
\dfrac{\partial \varphi}{\partial t}+(\mathbf v \nabla)\varphi(\mathbf
r, \mathbf v, t)=\dfrac{v}{l}\Big[2\beta\mathbf {v}
\mathbf {u_*}(\mathbf r, t)-\varphi(\mathbf r, \mathbf v,
t)\Big].
\eqno{(3.4)}
$$

Перейдём в уравнении (3.4) к безразмерной скорости молекул:
$\mathbf C=\sqrt{\beta}\mathbf{v}={\mathbf{v}}/{v_T}$.
Введём безразмерную (массовую) скорость газа
$
\mathbf {U}(\mathbf r,t)=\sqrt{\beta}\mathbf{u_*}(\mathbf r, t).
$

Тогда уравнение(3.4) записывается в следующем виде:
$$
\dfrac {\partial \varphi}{\partial t}+\mathbf v \dfrac{\partial
\varphi}{\partial \mathbf {r}}=\dfrac{v}{l}\Big[2 \mathbf
{C}\mathbf{U}(\mathbf r, t) - \varphi(\mathbf r, \mathbf v,
t)\Big].
$$

Введём новые безразмерные переменные:
$$
t_1=\dfrac{t}{\tau}=\nu_0 t, \quad, \nu_0=\dfrac{1}{\tau},\quad
\mathbf{r}_1=\dfrac{\mathbf{r}}{l}=\nu_0 \sqrt{\beta}\mathbf{r}.
$$

Тогда предыдущее уравнение в безразмерных переменных записывается в виде:
$$
\dfrac{\partial \varphi}{\partial t_1}+\mathbf{C}\dfrac{\partial
\varphi}{\partial \mathbf {r}_1}+C\varphi(\mathbf {r}_1, \mathbf
C, t_1)=2C \mathbf C \mathbf {U}(\mathbf {r}_1, t_1). \eqno
{(3.5)}
$$

Теперь воспользуемся законом сохранения импульса. В линейном
приближении этот закон принимает следующий вид:
$$
\int v \mathbf {v} (f-f^{*}_{F})d\Omega=0,
$$
или, в безразмерных переменных,
$$
\int C \mathbf{C}g(C)(2\mathbf{C\,U}_*(\mathbf
{r}_1,t_1)-\varphi)d^3C=0.
$$

Это векторное равенство эквивалентно трем скалярным, из которых
последовательно получаем:
$$\displaystyle
2 U_x^{*}=\dfrac{\displaystyle\int C C_x
\varphi(\mathbf{r_1},\mathbf {C},t_1)g(C)d^3C}{
\displaystyle\int C C_{x}^{2}g(C)d^3C},\;
2 U_{y}^{*}=\dfrac{\displaystyle\int C C_y
\varphi(\mathbf{r}_1,\mathbf
{C},t_1)g(C)d^3C}{\displaystyle\int C
C_{y}^{2}g(C)d^3C},
$$
$$
2 U_{z}^{*}=\dfrac{\displaystyle\int C C_z
\varphi(\mathbf{r}_1,\mathbf
{C},t_1)g(C)d^3C}{\displaystyle\int C
C_{z}^{2}g(C)d^3C}.
$$

Заметим, что в силу симметрии знаменатели всех этих трех
выражений равны друг другу. Вычислим их, перейдя к полярным координатам:
$
C_x=C\mu, \quad C_y=C \sin {\theta} \cos {\chi}, \quad
C_z=C\sin{\theta} \sin{\chi}.
$
Имеем:
$$
\int C
C_{y}^{2}g(C)d^3C=\int\limits_{-1}^{1}
\int\limits_{0}^{\infty}\int\limits_{0}^{2\pi}
CC^2(1-\mu^2)\cos^2{\chi}g(C)C^2d\mu dC d\chi=
$$
$$
=\dfrac{4\pi}{3}\int\limits_{0}^{\infty}C^5g(C)dC=
\dfrac{4\pi}{3}g_5(\alpha),
$$
где
$$
g_n(\alpha)=\int\limits_{0}^{\infty}
\dfrac{\exp(C^2-\alpha)C^5dC}{(1+\exp(C^2-\alpha))^2},\quad n=0,1,2,\cdots.
$$

Вычисляя по частям дважды интеграл из $g_5(\alpha)$, находим, что
$$
g_5(\alpha)=
2\int\limits_{0}^{\infty}\dfrac{C^3dC}{1+\exp(C^2-\alpha)}=
2\int\limits_{0}^{\infty}C\ln(1+\exp(\alpha-C^2))dC.
$$
Таким образов, мы нашли, что
$$
2\mathbf{U}^{*}(\mathbf {r}_1,t_1)=\dfrac{3}{4\pi g_5(\alpha)}\int C
\mathbf{C} \varphi(\mathbf{r}_1,\mathbf {C},t_1)g(C)d^3C.
$$
Итак, уравнение (3.5) записывается в следующем виде:
$$
\dfrac{\partial \varphi}{\partial t_1}+\mathbf{C}\dfrac{\partial
\varphi}{\partial {r}_1}+C\varphi(\mathbf{r}_1,\mathbf C,
t_1)=\hspace{5cm}$$$$=\dfrac{3C \mathbf{C}}{4\pi g_5(\alpha)}
\int C'\mathbf{C'}\varphi(\mathbf
{r}_1,\mathbf{C}',t_1)g(C')d^3C'.
%\eqno{(2.13)}
$$

В стационарных задачах предыдущее уравнение упрощается:
$$
\mathbf{C}\dfrac{\partial \varphi}{\partial {r}_1}+
C\varphi(\mathbf{r}_1,\mathbf C)=\dfrac{3C \mathbf{C}}{4\pi g_5(\alpha)}
\int C'\mathbf{C'}\varphi(\mathbf
{r}_1,\mathbf{C}',)g(C')d^3C'.
\eqno{(3.6)}
$$

Рассмотрим частный случай уравнения (3.6) для задач скольжения
квантового газа. Пусть газ заполняет полупространство
$\{x,y,z: x>0\}$ и движется вдоль плоской поверхности
в направлении оси $y$.

Тогда, согласно Черчиньяни \cite{105}, функцию распределения
будем искать в виде:
$$
\varphi(\mathbf {r_1}, \mathbf C)=C_y h(x_1,\mu,C).
$$

При этом мы предполагаем, что функция $h(x_1,\mu,C)$ не зависит от
азимутального угла в плоскости $(y,z)$, а зависит только от $\mu$, $C$ и
$x_1$ \;$(x_1=\mathbf{r}_{1x})$, где
$\mu=\dfrac{C_x}{C}$.
Учитывая, что в сферических координатах
$$
\mathbf{C\,C'}=C_x{C'}_x+C_y{C'}_y+C_z{C'}_z=
$$
$$
=C\mu C'\mu'+CC'\sqrt{1-\mu^2}\sqrt{1-{\mu'}^2}\Big(\cos \chi\cos \chi'+
\sin \chi\sin\chi'\Big),
$$
на основании (3.6) получаем следующее уравнение:
$$
\mu\dfrac{\partial h}
{\partial x_1}+h(x_1,\mu,C)=
$$
$$
=\dfrac{3}{4\pi g_5(\alpha)}\int C'
{C'_y}^2 h(x_1,\mu',C')g(C')d^3dC'.
\eqno{(3.7)}
$$

Упростим интегральное слагаемое из правой части уравнения
(3.7), переходя к сферическим координатам. Получаем, что
$$
\int C'
{C'_y}^2 h(x_1,\mu',C')g(C')d^3dC'=
$$
$$
=\int\limits_{-1}^{1}\int\limits_{0}^{\infty}\int\limits_{0}^{2\pi}
{C'}^5(1-\mu'^2)\cos^2{\chi}
h(x_1,\mu',C')g(C')d\mu'dC'd\chi'=
$$
$$
=\pi \int\limits_{-1}^{1}\int\limits_{0}^{\infty}
(1-\mu'^2)C'^5h(x_1,\mu',C')g(C')d\mu'dC'.
$$

Таким образом, уравнение (3.7) упрощается и имеет следующий
вид:
$$\begin{array}{c}
\mu \dfrac{ \partial h}{ \partial x_1}+ h(x_1, \mu, C)=
\\=\dfrac{3}{4} \displaystyle{} \int\limits_{-1}^{+1}(1-
\mu'^2)d \mu' \int\limits_{0}^{+ \infty} \exp(- C'^2) C'^5
 h(x_1, \mu', C')d C'.
 \end{array}
\eqno{(3.8)}
$$

%Заметим, что для квантовых бозе--газов полученное кинетическое
%уравнение будет точно таким же. Отличие состоит в том, что при
%выводе данного уравнения аналог максвелловской функции
%распределения заменяется на аналог бозевской функции распределения.

Из постановки задачи Крамерса будет видно, что граничные условия
в этой задаче не зависят от модуля молекулярной скорости, т.е.
$h(x_1,\mu,C) \equiv h(x_1,\mu)$. Тогда уравнение (3.8)
упрощается и имеет следующий вид
$$
\mu\dfrac{\partial h}{\partial x_1}+h(x_1,\mu)=
\dfrac{3}{4}\int\limits_{-1}^{1}(1-{\mu'}^2)h(x_1,\mu)d\mu.
\eqno{(3.9)}
$$

\begin{center}
  \item{}\section{Постановка задачи Крамерса с диффузными
граничными условиями}
\end{center}

\markboth{}{ПОСТАНОВКА ЗАДАЧИ КРАМЕРСА}

Задание градиента массовой скорости $g_v$ означает, что вдали от
стенки распределение массовой скорости в полупространстве имеет
линейный рост
$$
u_y(x)=u_{sl}+g_vx, \qquad x\to +\infty,
$$
где $u_{sl}$ -- неизвестная скорость скольжения.

Умножая это равенство на $\sqrt{\beta}$ и учитывая связь
размерного $g_v$ и безразмерного $G_v$ градиентов $g_v=\nu G_v$,
для безразмерной массовой скорости $U_y=\sqrt{\beta}u_y$ получаем
$$
U_y(x_1)=U_{sl}+G_vx_1,\qquad x_1\to +\infty.
\eqno{(4.1)}
$$

В соотношении (4.1) $U_{sl}=\sqrt{\beta}u_{sl}$ -- безразмерная
скорость скольжения, $x_1=\nu \sqrt{\beta}x=x/l$ --
безразмерная координата.

Диффузное отражение ферми--частиц от поверхности означает,
что последние отражаются от стенки, имея фермиевское
распределение,
т.е.
$$
f(x=0,\mathbf{v})=f_F(v),\qquad v_x>0.
\eqno{(4.2)}
$$

Учитывая, что функцию распределения для ферми--газов
мы ищем в виде
$$
f=f_F+g(C)C_yh(x_1,\mu),
$$
из условия (4.2)
получаем граничное условие на стенке на функцию $h(x_1,\mu)$:
$$
h(0,\mu)=0,\qquad \mu>0.
\eqno{(4.3)}
$$

Вторым граничным условием является граничное условие "вдали от
стенки"\,. Этим условием является соотношение (4.1).
Преобразуем это условие на функцию $h(x_1,\mu)$. Условие
(4.1) означает, что вдали от стенки массовая скорость переходит
в свое асимптотическое распределение
$$
U_y^{as}(x_1)=U_{sl}+G_vx_1.
$$
Выражение для массовой скорости (2.1) означает, что вдали от
стенки функция $h(x_1,\mu)$ переходит в свое асимптотическое
распределение
$$
h_{as}(x_1,\mu)=2U_{sl}(\alpha)+2G_v(x_1-\mu),
$$
называемое распределением Чепмена --- Энскога (см., например,
\cite{91}--\cite{95}).

Таким образом, вторым граничным условием является условие:
$$
h(x_1,\mu)=2U_{sl}(\alpha)+2G_v(x_1-\mu),\qquad x\to+\infty.
\eqno{(4.4)}
$$

Теперь задача Крамерса при условии диффузного отражения
фер\-ми--частиц от
стенки сформулирована полностью и состоит в
решении уравнения (3.9) с граничными условиями (4.3) и (4.4).
При этом требуется определить безразмерную скорость скольжения
$U_{sl}(\alpha)$, величина градиента $G_v$ считается заданной.
Кроме того, требуется построить функцию распределения в
полупространстве $x_1>0$, требуется построить профиль массовой
скорости $u_y(x_1)$ и найти значение массовой скорости газов
непосрественно у стенки $u_y(0)$.

%В заключение этой главы покажем, что параметр $u^*_y$ нелинейного
%кинетического уравнения (2.5) с массовой скоростью газа $u_y$.

Из предыдущих рассуждений по линеаризации уравнения (2.5)
вытекает, что $U_x^*(x_1)=U_z^*(x_1)=0$, а правая часть
уравнения (3.9) равна:
$$
2U_y^*(x_1)=\dfrac{3}{4}\int\limits_{-1}^{1}(1-{\mu'}^2)h(x_1,\mu')d\mu'.
\eqno{(4.5)}
$$

Возьмем теперь выражение для $y$--компоненты массовой скорости
газа
$$
u_y(x_1)=\int v_y f d\Omega \Big[\int fd\Omega\Big]^{-1}.
\eqno{(4.6)}
$$

После подстановки в (4.6) выражения $f=f_F+C_yg(C)h(x_1,\mu)$ для
ферми--газов и
последующей линеаризации получаем для безразмерной массовой
скорости следующее выражение:
$$
U_y(x_1)=\int C_y^2g(C,\alpha)h(x_1,\mu)d^3C\Big[\int f_Fd^3C\Big]^{-1}.
\eqno{(4.7)}
$$

Здесь
$$
\int f_Fd^3C=4\pi
\int\limits_{0}^{\infty}\dfrac{C^2dC}{1+\exp(C^2-\alpha)}=
2\pi l_0(\alpha),
$$
где
$$
l_0(\alpha)=\int\limits_{0}^{\infty}\ln(1+
\exp(\alpha-C^2))dC.
$$

Далее, замечая, что
$$
\int C_y^2g(C,\alpha)h(x_1,\mu)d^3C=\pi g_4(\alpha)
\int\limits_{-1}^{1}(1-\mu^2)h(x_1,\mu)d\mu,
$$
заключаем, что безразмерная массовая скорость равна:
$$
U_y(x_1)=\dfrac{g_4(\alpha)}{2l_0(\alpha)}
\int\limits_{-1}^{1}(1-\mu^2)h(x_1,\mu)d\mu=\dfrac{3}{8}
\int\limits_{-1}^{1}(1-\mu^2)h(x_1,\mu)d\mu.
\eqno{(4.8)}
$$
Теперь из (4.8) и (4.7) получаем, что $U_y$ и $U_y^*$ совпадают:
$$
U_y(x_1)=U_y^*(x_1).
\eqno{(4.9)}
$$

\begin{center}
\item{}\section{Собственные решения и дисперсионная функция задачи}
\end{center}
\markboth{}{СОБСТВЕННЫЕ РЕШЕНИЯ И ДИСПЕРСИОННАЯ ФУНКЦИЯ}
\begin{center}
  \item{} \subsection{Разделение переменных}
\end{center}

Разделение переменных в уравнении (3.9) осуществляется анзацем
Кейза
$$
h_ \eta(x, \mu)= \exp(- \dfrac{x}{ \eta}) \varphi( \eta, \mu),
\qquad \eta \in \overline{\mathbb{C}},
\eqno{(5.1)}
$$
где $\eta$ -- спектральный параметр, или параметр разделения,
вообще говоря, комплексный; $ \overline{\mathbb{C}}$ --
расширенная комплексная плоскость.

Если подставить выражение (5.1) в исходное уравнение (3.9), то
мы сразу получаем соответствующее характеристическое уравнение
$$
( \eta- \mu) \varphi( \eta, \mu)=
\dfrac{3}{4}\int\limits_{-1}^{1}
(1-{\mu'}^2)\varphi(\eta,\mu')\,d\mu'.
\eqno{(5.2)}
$$

Если ввести обозначение
$$
n(\eta)=\int\limits_{-1}^{1}
(1-{\mu'}^2)\varphi(\eta,\mu')\,d\mu',
\eqno{(5.3)}
$$
то характеристическое уравнение (5.2) можно переписать с помощью
(5.3) в виде
 $$
(\eta-\mu)\varphi(\eta,\mu)=\dfrac{3}{4}\eta n(\eta), \quad \eta
\in \overline{\mathbb{C}}.
\eqno{(5.4)}
$$
Условие (5.3) называется нормировочным условием,
нормировочным интегралом, или просто нормировкой.

В силу однородности исходного уравнения (3.9) можно считать, что
$$
n( \eta) \equiv \int\limits_{-1}^{1}
(1-{\mu'}^2)\varphi(\eta,\mu')\,d\mu'\equiv 1,
\eqno{(5.5)}
$$
т. е. считать, что нормировочный интеграл собственной функции
тождественно равен единице.

\begin{center}
 \item{}\subsection{Собственные функции и собственные
  значения}
\end{center}

Будем искать решение уравнений (5.4) и (5.5)
 в классе обобщенных функций $D'(\Delta),\; \Delta=(-1,1)$.

Общее решение этих уравнений в классе $D'(\Delta)$ дается (см.,
например, \cite{7}) формулой
$$
\varphi( \eta, \mu)= \dfrac{3}{4}\eta P \dfrac{1}{ \eta- \mu}+
g(\eta) \delta( \eta- \mu),
\eqno{(5.6)}
$$
когда $ \eta, \mu \in \Delta$.

Здесь $g( \eta)$ -- произвольная непрерывная функция,
определяемая ниже из условия нормировки (5.5),
%Подставляя (5.8) в (5.7),
%находим, что $g(\eta)=\lambda(\eta)/(5-\eta^2)$.
символ $Px ^{-1} $ означает распределение
  (обобщенную функцию) -- главное
значение интеграла по Коши при интегрировании выражения $x
^{-1}$, т. е. $Px ^{-1} $ означает линейный функционал,
действующий по формуле
$$
\left(P \dfrac{1}{x}, \varphi\right)= {\rm Vp}
\int\limits_{-1}^1 \dfrac{ \varphi(x)}{x}=\lim_{ \varepsilon \to
0} \left( \int\limits_{-1}^{- \varepsilon}+
\int\limits_{\varepsilon}^{+1}\right) \dfrac{
\varphi(x)}{x}\,dx,
$$
где $ \varphi\in D(\Delta)$.

Обобщенная функция $Px ^{-1}$ совпадает с обычной функцией $x
^{-1}$ при $x \in \Delta\setminus \{0\}$. Она называется главным
значением интеграла от $x ^{-1}$, $ \delta(x)$ есть
дельта--функция Дирака.

%Как уже отмечалось выше, функция $g( \eta)$ находится из условия (5.8).
Подставляя (5.6) в (5.5), получаем
$$
g(\eta)(1-\eta^2)=\lambda( \eta),
\eqno{(5.7)}
$$
где
$$
\lambda(z)=1+z \dfrac{3}{4}\int\limits_{-1}^{+1} (1- u^2)
\dfrac{du}{u-z}
\eqno{(5.8)}
$$
есть дисперсионная функция Вильямса \cite{61}, \cite{63},
\cite{68}, \cite{69}.

Подставляя $g( \eta)$ из выражения (5.7) в (5.6), находим:
$$
\varphi( \eta, \mu)= \dfrac{3}{4} \eta P \dfrac{1}{ \eta- \mu}+
\dfrac{\lambda(\eta)}{1-\eta^2} \delta(\eta- \mu),
\eqno{(5.9)}
$$
если $ \eta, \mu \in \Delta$.

Выражение (5.9) представляет собой собственную функцию
непрерывного спектра ($ \eta \in \Delta$), отвечающую единичной
нормировке (5.5).

Ясно, что непрерывный спектр характеристического уравнения
заполняет сплошным образом всю действительную ось, ибо при
изменении $ \mu$ непрерывным образом от $-1$ до $+1$ нуль
разности $ \eta- \mu$ пробегает непрерывным образом также всю
действительную ось
 $ \eta$, т. е.
$\eta \in (-1, +1)=\Delta$.

Дискретным спектром характеристического уравнения называют (см.,
например, \cite{14}, \cite{95}) множество нулей дисперсионного
уравнения
$$ \lambda(z)=0.
\eqno{(5.10)}
$$

Уравнение (5.10) $ \lambda(z)$ имеет единственный нуль --
двойной нуль в точке $ \eta_0= \infty$. Этому нулю, как кратной
точке дискретного спектра, отвечает два собственных решения
уравнения (3.9).

Одно из двух решений исходного уравнения (3.9) в силу его
однородности является произвольной постоянной. Второе
собственное решение уравнения (3.9) есть разность $x-\mu$:
$$
h_{1}(x, \mu)=1 \quad\mbox{и} \quad h_{2}(x, \mu)=x-\mu.
%\eqno{(3.14)}
$$
Эти решения будем называть решениями, отвечающими дискретному
спектру, или дискретными решениями уравнения (3.9). Эти решения
являются частными решениями уравнения (3.9). Следовательно, в
силу линейности уравнения (3.9) асимптотическая функция
распределения Чепмена --- Энскога $h_{as}(x,\mu)$ является его
решением.

Подставляя собственную функцию (5.9) в анзац Кейза, получаем
собственное решение исходного уравнения (3.9), отвечающие
непрерывному спектру
$$
h_ \eta(x, \mu)= \exp(-\dfrac{x}{\eta}) \left[ \dfrac{3}{4} \eta
P \dfrac{1}{ \eta- \mu}+ \dfrac{\lambda(\eta)}{1-\eta^2}
\delta(\eta- \mu) \right].
\eqno{(5.11)}
$$

Ниже будем говорить не об одной собственной функции (5.9)
(характеристического уравнения) непрерывного спектра,
 а о системе собственных
функций непрерывного спектра, имея ввиду то обстоятельство, что
каждому значению $ \eta \in \Delta$ отвечает конкретная
собственная функция. Следовательно, мы имеем континуальное
множество собственных функций (5.9) характеристического
уравнения и соответствующее семейство (5.11) собственных решений
кинетического уравнения (3.9).

\begin{center}
\item{} \subsection{ Дисперсионная функция и ее свойства}
\end{center}

Будем изучать основные свойства дисперсионной функции Вильямса
(5.8).
Нетрудно вычислить, что
$$
\lambda(z)=-\dfrac{1}{2}+ \dfrac{3}{2}(1-z^2)\lambda_c(z),
$$
где $\lambda_c(z)$-- дисперсионная функция Кейза \cite{14},
$$
\lambda_c(z)=1+\dfrac{1}{2}z\int\limits_{-1}^{1}
\dfrac{d\tau}{\tau-z}.
\eqno{(5.12)}
$$

Дисперсионная функция Вильямса $\lambda(z)$ аналитична во всей
комплексной плоскости за исключением точек разреза $\bar
\Delta=[-1,+1]$. Для граничных значений дисперсионной функции
Вильямса сверху и снизу на разрезе $\Delta=(-1,1)$ справедливы
формулы Сохоцкого
$$
\lambda^{\pm}(\mu)= \lambda(\mu)\pm i\dfrac{3}{4}\pi
\mu(1-\mu^2),
$$
где
$$
\lambda(\mu)= -\dfrac{1}{2}+\dfrac{3}{2}(1-\mu^2)\lambda_c(\mu),
$$
причем интеграл в выражении для $\lambda_c(\mu)$ понимается как
особый в смысле главного значения по Коши.

\begin{figure}
\begin{center}
\includegraphics{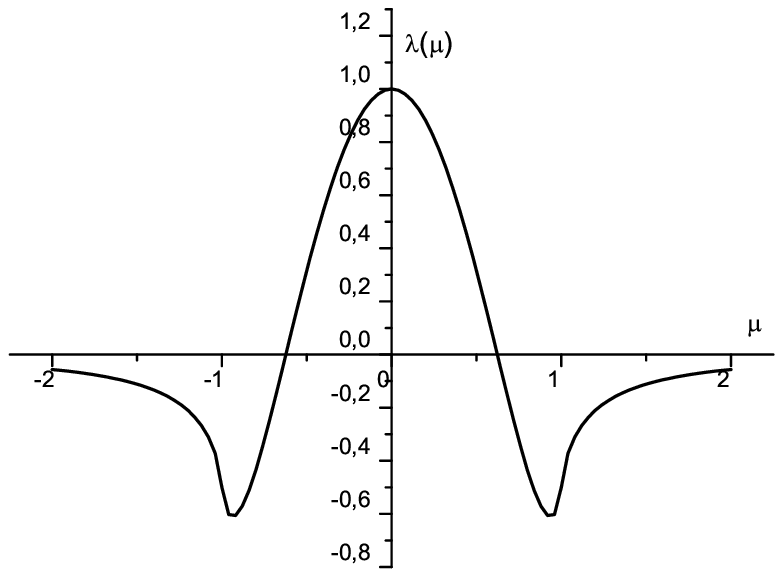}
\end{center}
\begin{center}
  Рис. 1. Действительная часть дисперсионной функция
  Вильямса на действительной оси.
\end{center}
\end{figure}%

Из формул (5.8) получаем
$$
\lambda^+(\mu)-\lambda^-(\mu)= i\pi\frac{3}{2}\mu(1-\mu^2),
\qquad \mu\in \Delta,
%\eqno{(5.4)}
$$
$$
\frac{1}{2}[\lambda^+(\mu)+ \lambda^-(\mu)]=\lambda(\mu), \qquad
\mu\in \Delta.
$$
Здесь $\lambda(\mu)$ на действительной оси выражается равенством
$$
\lambda(\mu)=1+\mu\frac{3}{4}\int\limits_{-1}^{1}
\dfrac{1-\tau^2}{\tau-\mu} \,d\tau=
%$$$$=
-\dfrac{1}{2}+\dfrac{3}{2}(1-\mu^2)
\left(1+\dfrac{\mu}{2}\ln\dfrac{1-\mu}{1+\mu}\right).
$$

\begin{figure}
\begin{center}
\includegraphics{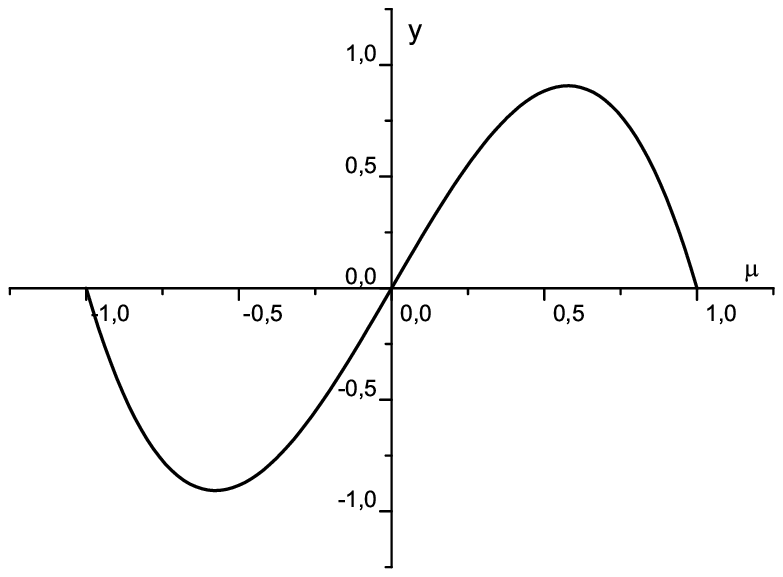}
\end{center}
\begin{center}
  Рис. 2. Мнимая часть дисперсионной функции
   Вильямса на отрезке $[-1,1]$ действительной оси.
\end{center}
\end{figure}

Разложим функцию $\lambda(z)$ в ряд Лорана. Воспользуемся
формулой (5.12). Сначала разложим в ряд Лорана дисперсионную
функцию Кейза при $|z|>1$:
$$
\lambda_c(z)=-\dfrac{1}{3z^2}-\dfrac{1}{5z^4}-
\dfrac{1}{7z^6}-\cdots-\dfrac{1}{(2n+1)z^{2n}}- \cdots.
\eqno{(5.13)}
$$

Подставляя разложение (5.13) в (5.12), находим разложение для
дисперсионной функции Вильямса при $|z|>1$:
$$
\lambda(z)=-\dfrac{1}{5z^2}-
\dfrac{3}{35z^4}-\dfrac{1}{21z^6}-\cdots-
\dfrac{3}{(2n+1)(2n+3)z^{2n}}-\cdots .
\eqno{(5.14)}
$$
Из разложения (5.14) видно, что точка $z=\infty$ является
двойным нулем функции $\lambda(z)$. Этому нулю, как двойной
точке дискретного спектра, отвечают два собственных решения
уравнения (3.9) $h_1(x,\mu)$ и $h_2(x,\mu)$.

%В верхней (нижней) полуплоскости функция $\lambda(z)$ вычисляется по формуле,
%аналогичной (5.1):

Из равенств (5.8), (5.12) и формул Сохоцкого для $\lambda(z)$
видно, что
$$
\begin{array}{l}
\Re \lambda^{\pm}(\mu)=\lambda(\mu),
\quad \mu\in \Delta, \\
\Im \lambda^{\pm}(\mu)=\pm \pi\frac{3}{4}\mu(1-\mu^2), \quad \mu
\in \Delta.
\end{array}
$$

Приведем графики функций $y=\lambda(\mu),\;\mu\in \Delta$ и
$y=\pi\frac{3}{4}\mu(1-\mu^2),\;\mu\in \Delta$ (см. рис. 1 и
рис. 2). Построим на комплексной плоскости (см. рис. 3)
кривые $\Gamma^{\pm}:z=\lambda^{\pm}(\mu),\;\mu\in \Delta$,
которые определяются следующими параметрическими уравнениями:
$$ \Gamma^{\pm}:\left\{
\begin{array}{l}
x=\Re \lambda(\mu), \\ y=\pm \Im \lambda(\mu),
\end{array}
\right. \mu\in \Delta.
$$

Эти кривые представляют собой одно и то же геометрическое место
точек $\Gamma$ (см. рис. 3), причем кривая $\Gamma^+$
ориентирована положительно, а $\Gamma^-$ -- отрицательно.
\begin{figure}
\begin{center}
\includegraphics[width=14.0cm, height=8cm]{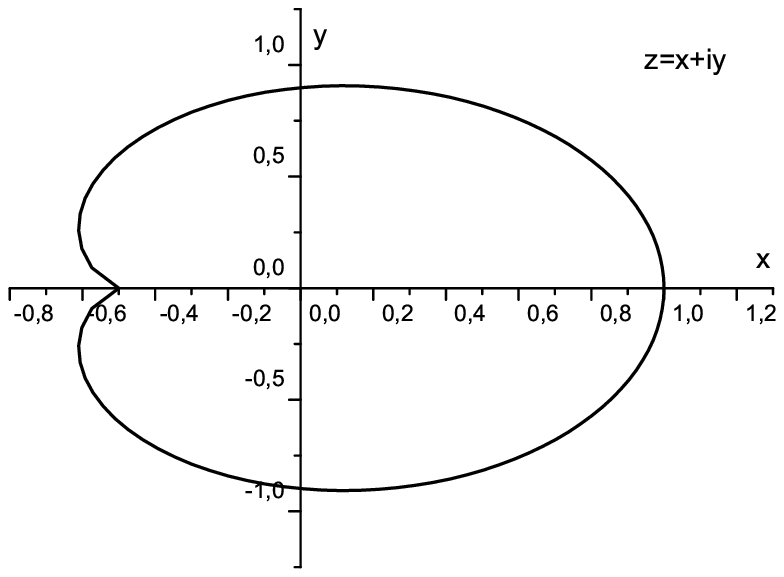}
\end{center}
\begin{center}
  Рис. 3. Кривая $\Gamma: z=\lambda^+(\mu),\;-1<\mu<1,$
определяет замкнутую кривую на комплексной плоскости.
\end{center}
\end{figure}

Введем угол $\theta(\mu)=\arg \lambda^+(\mu)$ -- главное
значение аргумента функции $\lambda^+(\mu)$, фиксированное в
нуле условием $\theta(0)=0$. Из рис. 3 видно, что приращение
этого угла на отрезке $[0,1]$ равно $\pi$, а на отрезке $[-1,1]$
равно $2\pi$. Угол $\theta(\mu)$ вычисляется по формуле:
$$
 \theta(\mu)=\arcctg \dfrac{4\lambda(\mu)}
 {3\pi \mu(1-\mu^2)}.
$$

\begin{figure}
\begin{center}
\includegraphics{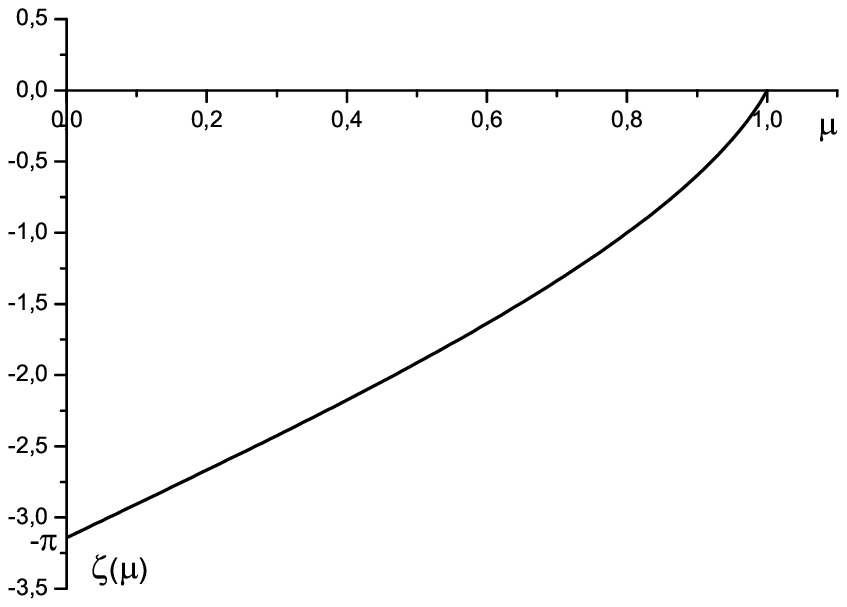}
\end{center}
\begin{center}
  Рис. 4. Кривая $\zeta=\zeta(\mu),\quad 0<\mu<1.$
\end{center}
\end{figure}

Введем еще один угол $\zeta(\mu)=\theta(\mu)-\pi$ (см. рис.
4), который понадобится ниже. Его удобно вычислять по формуле:
$$
%\begin{array}{l}
\zeta(\mu)=\arcctg \dfrac{4\lambda(\mu)} {3\pi
\mu(1-\mu^2)}-\pi=% $$$$=
-\dfrac{\pi}{2}-\arctg
\dfrac{4\lambda(\mu)} {3\pi \mu(1-\mu^2)}.
%\end{array}
$$

Из формул Сохоцкого для $\lambda(z)$ следует также, что
$$
|\lambda^{\pm}(\mu)|^2=\lambda^2(\mu)+\Big(\pi\frac{3}{4}
\mu(1-\mu^2)\Big)^2= \lambda^+(\mu)\lambda^-(\mu), \quad \mu\in
\Delta.
$$

\begin{figure}
\begin{center}
\includegraphics{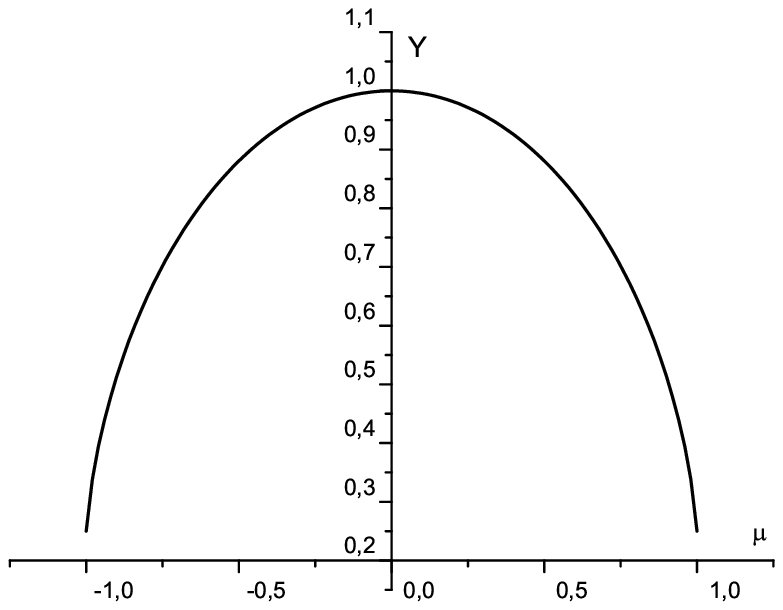}
\end{center}
\begin{center}
 Рис. 5. Кривая $Y=\lambda^+(\mu)\lambda^-(\mu),\; -1<\mu<1.$
\end{center}
\end{figure}

Функции $\lambda^+(\mu)$ и $\lambda^-(\mu)$ являются
комплексно--соп\-ря\-жен\-ны\-ми, т.~e.
$
\overline{\lambda^+(\mu)}= \lambda^-(\mu)$, $
\overline{\lambda^-(\mu)}=\lambda^+(\mu),
$
поэтому
$$
\lambda^+(\mu)=|\lambda^+(\mu)|\exp({\theta(\mu)}), \; \qquad
\lambda^-(\mu)=|\lambda^+(\mu)|\exp({-\theta(\mu)}),\;
$$$$
|\lambda^+(\mu)|=|\lambda^-(\mu)|,\; \quad\mu\in \Delta.
$$

Отсюда следует, что
$$
G(\mu)\equiv \dfrac{\lambda^+(\mu)}{\lambda^-(\mu)}=
\exp({2\theta(\mu)}), \quad |G(\mu)|\equiv 1, \quad \mu\in
\Delta.
$$

 Это значит, что кривая $\gamma:z=G(\mu), \;
\mu\in\Delta=(-1,+1)$, лежит на окружности единичного радиуса и
дважды обходит эту окружность в положительном направлении; если
же $\mu\in\Delta^+=(0,1)$, то эта кривая один раз обходит
единичную окружность в положительном направлении, это значит,
что ее индекс равен единице.

Введем еще одну кривую $\gamma^+:z=\lambda^+(\mu),
\mu\in\Delta^+$. Эта кривая совпадает (см. рис. 3) с частью
кривой $\Gamma^+$, лежащей в верхней полуплоскости.

Покажем, что $\lambda(z)$ не имеет конечных комплексных нулей в
области $D=\mathbb{C}\setminus [-1,+1]$. Возьмем контур
$\gamma_\varepsilon$, охватывающий отрезок
$[-1,+1]$ и отстоящий от него на расстоянии $\varepsilon,
\varepsilon>0$. Согласно принципу аргумента \cite{13},
\cite{14}, \cite{93} разность между числом нулей и полюсов в
области $D$ равна индексу кривой $\Gamma$, т. е. индексу
коэффициента $G(\mu)$:
$$
N-P=\dfrac{1}{2\pi i} \int\limits_{\gamma_\varepsilon}d\ln
\lambda(z)= \dfrac{1}{2\pi}\int\limits_{\gamma_\varepsilon}\,
d\,\arg \lambda(z),
$$
причем каждый нуль или полюс считаются столько раз, какова их
кратность.

Переходя к пределу при $\varepsilon\to 0$ в этом равенстве,
получаем:
$$
\begin{array}{c}\displaystyle
N-P=\dfrac{1}{2\pi}\int\limits_{-1}^{1} d\,\arg
\dfrac{\lambda^+(\mu)} {\lambda^-(\mu)}=\dfrac{1}{2
\pi}\int\limits_{-1}^{1} d\,\arg G(\mu)= \\\displaystyle
=\dfrac{1}{\pi}\int\limits_{-1}^{1}\,d\,\theta(\mu)=
\dfrac{\theta(1)-
 \theta(-1)}{\pi}=\dfrac{\pi-(-\pi)}{\pi}=2.
\end{array}
$$
Отсюда следует, что $N=2$, ибо $P=0$ (полюсы у $\lambda(z)$
отсутствуют). Значит, других нулей, кроме двойного нуля
$z=\infty$, у функции $\lambda(z)$ в области $D$ нет.

\begin{center}
\item\section{Однородная краевая задача Римана}
\end{center}

\markboth{}{КРАЕВАЯ ЗАДАЧА РИМАНА}
\begin{center}
  \item{}\subsection{Факторизующая функция $X(z)$}
\end{center}

В основе аналитического решения граничных задач кинетической
теории лежит решение однородной краевой задачи Римана с
коэффициентом $G(\mu)=\lambda^+(\mu)/\lambda^-(\mu)$:
$$
\dfrac{X^+( \mu)}{X^-( \mu)}= \dfrac{ \lambda^+( \mu)}{
\lambda^-( \mu)}, \qquad 0<\mu<1. \eqno{(6.1)}
$$

Задача факторизации (6.1) для коэффициента задачи Римана
называется также ( см. \cite{13}, \cite{14}) однородной краевой
задачей Римана.

Учитывая свойства функций $ \lambda^{\pm}( \mu)$, перепишем
краевое условие (6.1) в виде
$$
\dfrac{X^+( \mu)}{X^-( \mu)}= \exp [2i( \theta( \mu)+ \pi k)],\;
k=0,\pm 1, \pm 2,\cdots , \; 0<\mu<1.
$$

Логарифмируя это краевое условие, получаем
$$
\ln X^+( \mu)-\ln X^-( \mu)= 2 i ( \theta( \mu)+ \pi k), \;
k=0,\pm 1,\pm 2,\cdots , \; 0<\mu<1.
$$
Решение этой задачи, как задачи определения аналитической
функции по скачку, имеет вид
$$
\ln X(z)= \dfrac{1}{ \pi} \int\limits_{0}^{1} \dfrac{ \theta(
\mu)+ \pi k}{ \mu-z}\,d \mu,\; k=0,\pm 1, \pm 2, \cdots .
\eqno{(6.2)}
$$

Обозначим
$$
V(z)= \dfrac{1}{ \pi} \int\limits_{0}^{1} \dfrac{ \theta( u)-
\pi}{u-z}\,du.
\eqno{(6.3)}
$$

\begin{figure}
\begin{center}
\includegraphics{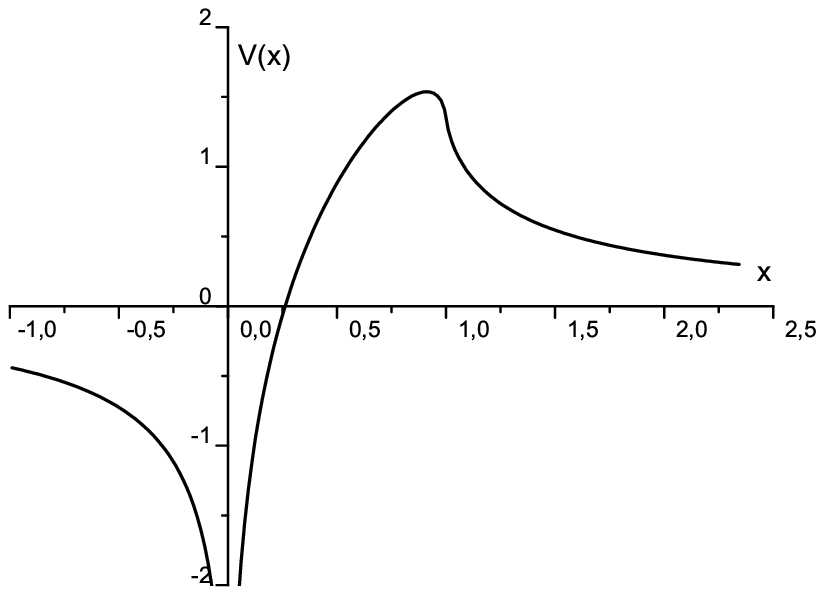}
\end{center}
\begin{center}
  Рис. 6. Функция $V=V(x)$ на действительной оси.
\end{center}

\end{figure}

Тогда из равенства (6.2) имеем:
$$
X(z)= \exp V(z).
\eqno{(6.4)}
$$

Выясним поведение функции $X(z)$ в начале координат. При $z \to
0$ из формулы (6.3) имеем:
$$
V(z)=- \dfrac{ \theta(0)- \pi}{ \pi}\ln z +O_1(z) \quad (z \to
0),
$$
где $ \theta(0)=0$, а $O_1(z)$ -- ограниченная функция в начале
координат.

Теперь видно, что функция $X(z)=z \exp O_1(z)$ является
исчезающей в начале координат.

Чтобы сделать решение задачи (6.1) неисчезающим в нуле,
переопределим решение (6.4) следующим образом:
$$
X(z)= \dfrac{1}{z} \exp V(z).
\eqno{(6.5)}
$$

Эту функцию далее будем называть факторизующей, ибо она
осуществляет факторизацию коэффициента задачи $ G( \mu)$.

Учитывая, что $\zeta(\tau)=\theta(\tau)-\pi$, решению (6.5)
можно придать следующую форму:
$$
X(z)=\dfrac{1}{z-1}\exp V^\circ(z), \eqno{(6.6)}
$$
где
$$
V^\circ(z)= \dfrac{1}{\pi}
\int\limits_{0}^{1}\dfrac{\theta(\tau)\,d\tau}{\tau-z}.
$$

\begin{figure}
\begin{center}
\includegraphics{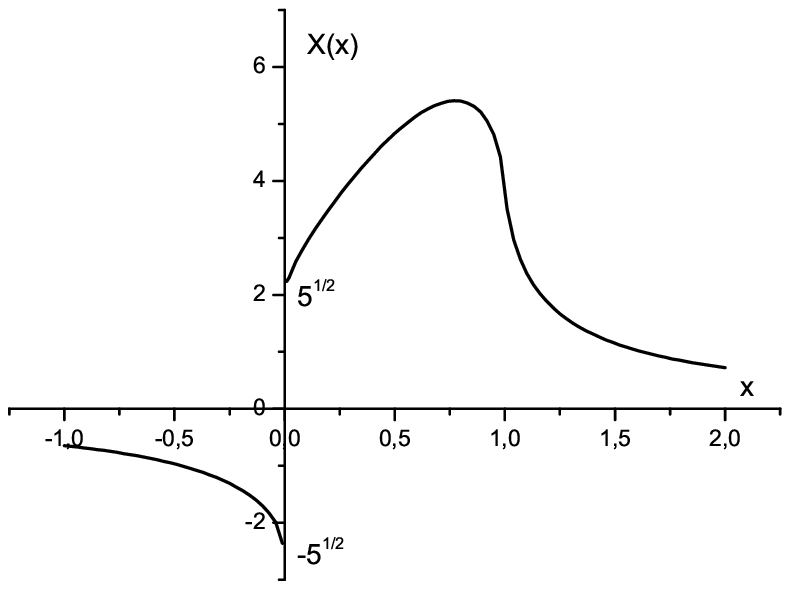}
\end{center}

\begin{center}
  Рис. 7. Функция $X=X(x)$ на действительной оси.
\end{center}

\end{figure}

В окрестности начала координат имеем:
$$
X(z)=\dfrac{1}{z-1}\exp\left[-\dfrac{\theta(0)}{\pi} \ln
z+O_1(z)\right]= \dfrac{1}{z-1}\exp O_1(z)
$$
-- ограниченная функция, в окрестности точки $z=1$ имеем:
$$
X(z)=\dfrac{1}{z-1}\exp\left[\dfrac{\theta(1)}{\pi} \ln
(z-1)+O_2(z)\right]= \exp O_2(z)
$$
-- ограниченная функция, ибо $O_2(z)$ -- ограниченная в
окрестности точки $z=1$ функция.

Итак, решение (6.6) задачи (6.1) есть функция, ограниченная в
окрестности концевых точек отрезка интегрирования.

\begin{center}
\item{}\subsection{Интегральные представления
функций $X(z)$ и $X^{-1}(z)$}
\end{center}

Возьмем замкнутый контур,  состоящий из
окружности $\Gamma_R$ радиуса $R=1/\varepsilon$ и контура
$\gamma_\varepsilon$, охватывающего разрез $\bar \Delta^+=[0,1]$
по часовой стрелке и отстоящего от разреза на расстоянии
$\varepsilon$. Радиус малых полуокружностей, входящих в
$\gamma_\varepsilon$, равен $r=\varepsilon$. Пусть $D_R$ --
Двухсвязная область с границей $\partial D_R=\Gamma_R\cup
\gamma_\varepsilon$.

Возьмем произвольную точку $z\in \mathbb{C}\setminus [0,1]$.
Число $\varepsilon$ выберем столь малым, чтобы $z\in D_R$.
Согласно интегральной формуле Коши для многосвязных областей
$$
X(z)=\dfrac{1}{2\pi i}\left(\int\limits_{\Gamma_R}+
\int\limits_{\gamma_\varepsilon}\right)
\dfrac{X(z')\,dz'}{z'-z}.
$$
Переходя к пределу при $\varepsilon\to 0$ в последнем равенстве,
получаем следующее интегральное представление
$$
X(z)=\dfrac{1}{2\pi i}\int\limits_{0}^{1}
\dfrac{X^+(\tau)-X^-(\tau)} {\tau-z}d\tau, \quad z\in
\mathbb{C}\setminus [0,1].
\eqno{(6.7)}
 $$

Интегральное представление (6.7) выражает значение функции
$X(z)$ в каждой точке $z\in\mathbb{C}\setminus [0,1]$ через
разность граничных значений этой функции сверху и снизу на
разрезе $[0,1]$.

Используя задачу Римана (6.1) плотность интегрального
представления (6.7) можно представить в двух эквивалентных
выражениях:
$$
X^+(\tau)-X^-(\tau)=
X^+(\tau)-\dfrac{\lambda^-(\tau)}{\lambda^+(\tau)} X^+(\tau)=
$$
$$
= \dfrac{3\pi i}{2}\tau(1-\tau^2)\dfrac{X^+(\tau)}
{\lambda^+(\tau)},
$$
$$
X^+(\tau)-X^-(\tau)=2i\sin \zeta(\tau),\quad \tau\in (0,1).
\eqno{(6.8)}
$$
Следовательно, на основании (6.7) получаются следующие два
интегральных представления:
$$
X(z)=\dfrac{3}{4}\int\limits_{0}^{1}
\dfrac{X^+(\tau)}{\lambda^+(\tau)}
\dfrac{\tau(1-\tau^2)}{\tau-z}d\tau,\quad
z\in\mathbb{C}\setminus [0,1], \eqno{(6.9)}
$$
$$
X(z)=\dfrac{1}{\pi}\int\limits_{0}^{1}
X(\tau)\dfrac{\sin\zeta(\tau)\,d\tau}{\tau-z},\quad
z\in\mathbb{C}\setminus [0,1]. \eqno{(6.10)}
$$

Интегральное представление (6.9) выражает значение функции
$X(z)$ в каждой точке $z\in\mathbb{C}\setminus [0,1]$ через свои
же значения на разрезе, взятые "сверху".

Интегральное представление (6.10) выражает значение функции
$X(z)$ в каждой точке $z\in\mathbb{C}\setminus [0,1]$ через свои
значения на разрезе.

Аналогично интегральному представлению (6.7) для функции \;
$X^{-1}(z)$ можно вывести следующее интегральное представление

$$
\dfrac{1}{X(z)}-z+V_1=\dfrac{1}{2\pi i}\int\limits_{0}^{1}
\left[\dfrac{1}{X^+(\tau)}-
\dfrac{1}{X^-(\tau)}\right]\dfrac{d\tau} {\tau-z}, $$$$ z\in
\mathbb{C}\setminus [0,1], \eqno{(6.11)}
$$

Интегральное представление (6.11) восстанавливает значение
функции $X^{-1}(z)$ в точке $z\in \mathbb{C}\setminus [0,1]$
через разность своих граничных значений на разрезе. Здесь
$$
V_1=-\dfrac{1}{\pi}\int\limits_{0}^{1}\zeta(\tau)\,d\tau.
$$

Замечая, что согласно (6.1), (6.3) и (6.5)
$$
\dfrac{1}{X^+(\tau)}-\dfrac{1}{X^-(\tau)}=- \dfrac{3}{2}\pi
i\dfrac{\tau(1-\tau^2)} {X^+(\tau)\lambda^-(\tau)},
$$
$$
\dfrac{1}{X^+(\tau)}-\dfrac{1}{X^-(\tau)}=-
2i\dfrac{\sin\zeta(\tau)}{X(\tau)}, \eqno{(6.12)}
$$
получаем следующее интегральное представление:
$$
\dfrac{1}{X(z)}=z-V_1-\dfrac{1}{\pi}\int\limits_{0}^{1}
\dfrac{\sin\zeta(\tau)}{X(\tau)} \dfrac{d\tau}{\tau-z},\quad
z\in \mathbb{C}\setminus [0,1]. \eqno{(6.13)}
$$

Выведем интегральные представления функций $X(z)$ и $X^{-1}(z)$
для точек разреза, т. е. когда $z=\mu\in(0,1)$.

Для точек $\mu\in (0,1)$ согласно формуле Сохоцкого на основании
(6.7) имеем
$$
X^+(\mu)=\dfrac{1}{2}[X^+(\mu)-X^-(\mu)]+\dfrac{1}{2\pi i}
\int\limits_{0}^{1}\dfrac{X^+(\tau)-X^-(\tau)} {\tau-\mu}d\tau,
%\eqno{(4.2)}
$$
откуда
$$
\dfrac{X^+(\mu)+X^-(\mu)}{2}=\dfrac{1}{2\pi i}
\int\limits_{0}^{1}
\dfrac{X^+(\tau)-X^-(\tau)}{\tau-\mu}d\tau,\quad \mu\in (0,1).
%\eqno{(6.14)}
$$

Таким образом, сумма граничных значений функции $X(z)$ на
разрезе выражается через разность своих граничных значений
на этом разрезе.% с помощью представления (6.14).

Аналогично можно получить, что при $\mu\in(0,1)$
$$
\dfrac{1}{2}\left[\dfrac{1}
{X^+(\mu)}+\dfrac{1}{X^-(\mu)}\right]=\mu-V_1+ \dfrac{1}{2\pi
i}\int\limits_{0}^{1} \left[\dfrac{1}{X^+(\tau)}-
\dfrac{1}{X^-(\tau)}\right]\dfrac{d\tau} {\tau-\mu}.
$$

Теперь учитывая, что
$$
\dfrac{X^+(\mu)+X^-(\mu)}{2}=X(\mu)\cos \zeta(\mu),
$$
$$
\dfrac{1}{2}\left[\dfrac{1}
{X^+(\mu)}+\dfrac{1}{X^-(\mu)}\right]=\dfrac{\cos\zeta(\mu)}
{X(\mu)},
$$
получаем следующие два интегральных представления
$$
X(\mu)\cos \zeta(\mu)=\dfrac{1}{2\pi i}
\int\limits_{0}^{1}\dfrac{X^+(\tau)-X^-(\tau)} {\tau-\mu}d\tau,
\quad \mu\in(0,1),
%\eqno{(6.10)}
$$
$$
\dfrac{\cos \zeta(\mu)}{X(\mu)}=\mu-V_1+ \dfrac{1}{2\pi
i}\int\limits_{0}^{1} \left[\dfrac{1}{X^+(\tau)}-
\dfrac{1}{X^-(\tau)}\right]\dfrac{d\tau} {\tau-\mu}, \quad
\mu\in(0,1),
%\eqno{(6.12)}
$$
или, учитывая равенства (6.8) и (6.12), отсюда получаем:
$$
X(\mu)\cos \zeta(\mu)=\dfrac{1}{\pi}
\int\limits_{0}^{1}X(\tau)\dfrac{\sin\zeta(\tau)}
{\tau-\mu}d\tau, \quad \mu\in(0,1),
%\eqno{(6.10)}
$$
$$
\dfrac{\cos \zeta(\mu)}{X(\mu)}=\mu-V_1-
\dfrac{1}{\pi}\int\limits_{0}^{1}
\dfrac{\sin\zeta(\tau)}{X(\tau)}\dfrac{d\tau} {\tau-\mu}, \quad
\mu\in(0,1).
%\eqno{(6.12)}
$$

Последние четыре интегральных представления можно распространить
с разреза на всю комплексную плоскость. Для этого заменить угол
$\zeta(\mu)$ в левых частях интегральных представлений на угол
$$
\zeta_+(\mu)=\left\{
\begin{array}{l}
\zeta(\mu), \quad 0\le \mu<1, \\0, \qquad \mu\notin [0,1].
\end{array}
\right.%\}
$$

В частности, для точек разреза дисперсионной функции $(-1,1)$
имеем:
$$
X(\mu)\cos \zeta_+(\mu)=\dfrac{1}{\pi}
\int\limits_{0}^{1}X(\tau)\dfrac{\sin\zeta(\tau)}
{\tau-\mu}d\tau, \quad \mu\in(-1,1),
%\eqno{(6.14)}
$$
$$
\dfrac{\cos \zeta_+(\mu)}{X(\mu)}=\mu-V_1-
\dfrac{1}{\pi}\int\limits_{0}^{1}
\dfrac{\sin\zeta(\tau)}{X(\tau)}\dfrac{d\tau} {\tau-\mu}, \quad
\mu\in(-1,1).
%\eqno{(6.15)}
$$

Здесь
$$
X(\mu)=\dfrac{1}{\mu}\exp V(\mu), \quad
V(\mu)=\dfrac{1}{\pi}\int\limits_{0}^{1}
\dfrac{\zeta(\tau)d\tau}{\tau-\mu},
$$
$$
\zeta_+(\mu)=\left\{
\begin{array}{l}
\zeta(\mu), \quad 0\le \mu<1, \\0, \qquad -1<\mu<0,
\end{array}
\right\}=\zeta(\mu)H_+(\mu).
$$

Последние два интегральных представления будут использованы ниже
при построении функции распределения для $\mu\in (-1,1)$.

\begin{center}
\item{}\subsection{Факторизация дисперсионной функции}
\end{center}

Покажем, что для дисперсионной функции $\lambda(z)$ везде в
комплексной плоскости, за исключением точек разреза $[0,1]$,
справедлива формула
$$
\lambda(z)=\dfrac{1}{5}X(z)X(-z), \qquad z\in
\mathbb{C}\setminus [-1,1].
\eqno{(6.14)}
$$

Из формулы (6.14) вытекает, что для граничных значений
дисперсионной функции на разрезе справедливы соотношения:
$$
\lambda^{\pm}(\mu)=\dfrac{1}{5}X^{\pm}(\mu)X(-\mu), \qquad 0\le
\mu<1,
%\eqno{(6.15)}
$$
$$
\lambda^{\pm}(\mu)=\dfrac{1}{5}X(\mu)X^{\mp}(-\mu), \qquad
-1<\mu<0.
$$

Для доказательства формулы (6.14) введем вспомогательную функцию
$$
R(z)=\dfrac{5\lambda(z)}{X(z)X(-z)}.
$$
Эта функция аналитична везде в комплексной плоскости, кроме
точек разреза $[-1,1]$. Каждая точка этого разреза является
устранимой. В самом деле, пусть $0\le \mu\le 1$. На основании
краевого условия (6.1) имеем:
$$
\dfrac{\lambda^+(\mu)}{X^+(\mu)X(-\mu)}=
\dfrac{\lambda^-(\mu)}{X^-(\mu)X(-\mu)},
$$
откуда $R^+(\mu)=R^-(\mu),\;0\le \mu \le 1$.

Пусть теперь $-1\le \mu<0$. Заменим $\mu$ на $-\mu$ в равенстве
(6.1):
$$
\dfrac{X^+(-\mu)}{X^-(-\mu)}=\dfrac{\lambda^+(-\mu)}
{\lambda^-(-\mu)}.
$$
Не трудно видеть, что $\lambda^+(-\mu)=\lambda^-(\mu)$,
$\lambda^-(-\mu)=\lambda^+(\mu)$. Поэтому, предыдущее равенство
можно переписать в виде
$$
\dfrac{X^+(-\mu)}{X^-(-\mu)}=\dfrac{\lambda^-(\mu)}
{\lambda^+(\mu)},
$$
отсюда
$$
\dfrac{\lambda^+(\mu)}{X(\mu)X^-(-\mu)}= \dfrac{\lambda^-(\mu)}
{X(\mu)X^+(-\mu)}.
$$
Следовательно, при $-1\le\mu<0\; R^+(\mu)=R^-(\mu)$. Таким
образом, функцию $R(z)$ можно считать аналитической функцией
везде в комплексной плоскости, в том числе и в точках разреза,
доопределив ее на разрезе по непрерывности. Осталось заметить,
что функция $R(z)$ аналитична и в бесконечно удаленной точке,
причем $R(\infty)=1$. По теореме Лиувилля эта функция является
тождественно постоянной: $R(z)\equiv 1$, откуда и вытекает
формула (6.14).

Из формулы (6.14) видно, что $X^2(0)=5$, откуда получаем:
$$
X(-0)=-\sqrt{5},\;\qquad X(+0)=\sqrt{5}.
$$

\begin{center}
\item{}\subsection{Вычисление коэффициентов
разложений функций $V(z)$, $X(z)$ и $X^{-1}(z)$}
\end{center}

В отличие от соответствующих разложений в случае постоянной
частоты столкновений молекул здесь все рассматриваемые ряды
являются сходящимися степенными рядами, а не асимптотическими,
как это было в случае постоянной частоты столкновений молекул.

Представим функцию $V(z)$ в окрестности точки $z=\infty$
сходящимся рядом Лорана
$$
V(z)=\dfrac{V_1}{z}+\dfrac{V_2}{z^2}+ \dfrac{V_3}{z^3}+\cdots,
\quad |z|>1. \eqno{(6.15)}
$$
Здесь
$$
V_n=-\dfrac{1}{\pi}\int\limits_{0}^{1}
\zeta(\tau)\tau^{n-1}\,d\tau, \quad n=1,2,3,\cdots .
$$

Выпишем численные значения нескольких первых коэффициентов
разложения (6.15):
$$
\begin{array}{l}
V_1=0.582,\quad V_2=0.214,\quad V_3=0.115,\\
V_4=0.073,\quad V_5=0.051,\quad V_6=0.038, \cdots .
\end{array}
$$

Можно показать, что четные коэффициенты находятся без квадратур
$$\extrarowheight=10pt
\begin{array}{l}
V_2=\dfrac{3}{14}=0.214\cdots, \qquad
V_4=\dfrac{1}{8}\left(\dfrac{20}{21}-
\dfrac{12}{7}V_2\right)=\dfrac{43}{588}=0.073\cdots,\\
V_6=\dfrac{1}{6}\left(\dfrac{5}{11}-\dfrac{10}{21}V_2-
\dfrac{12}{7}V_4\right)=\dfrac{857}{22638}=0.038\cdots,
\end{array}
$$

Найдем коэффициенты разложения функции $X(z)$ в окрестности
точки $z=\infty$. Пусть
$$
X(z)=\dfrac{1}{z}\exp(V(z)=\dfrac{1}{z}+\dfrac{X_2}{z^2}+
\dfrac{X_3}{z^3}+\dots, \qquad z \to \infty. \eqno{(6.16)}
$$
Отсюда
$$
\exp(V(z))=1+\dfrac{X_2}{z}+\dfrac{X_3}{z^2}+\dfrac{X_4}{z^3}+
\dots, \qquad z \to \infty.
$$
Логарифмируя это равенство и подставляя вместо функции $V(z)$ ее
разложение (6.15), получаем:
$$
\dfrac{V_1}{z}+\dfrac{V_2}{z^2}+\dots=\ln\left(1+
\dfrac{X_2}{z}+\dfrac{X_3}{z^2}+ \dfrac{X_4}{z^3}+\dots\right).
$$
Дифференцируя это равенство, получаем:
$$
\Big(\dfrac{V_1}{z^2}+\dfrac{2V_2}{z^3}+\dfrac{3V_3}{z^4}+
\dots\Big)\Big(1+\dfrac{X_2}{z}+\dfrac{X_3}{z^2}+
\dots\Big)=\dfrac{X_2}{z^2}+\dfrac{2X_3}{z^3}+\dfrac{3X_4}
{z^4}+\dots .
$$
Приравнивая коэффициенты при одинаковых степенях $z$, находим,
что коэффициенты разложения (6.16) вычисляются согласно
равенствам
$$%\extrarowheight=10pt
\begin{array}{l}
X_2=V_1=0.582,\qquad X_3=V_2+\dfrac{1}{2}V_1^2=0.384,\cdots.\\
% X_3=V_2+\dfrac{1}{2}V_1^2=0.384,\\
%X_4=V_3+V_1V_2+\dfrac{1}{6}V_1^3=0.273, \\
%X_5=V_4+V_1V_3+\dfrac{1}{2}V_2^2+\dfrac{1}{2}V_1^2V_2+\dfrac{1}{24}V_1^4=0.204.
\end{array}
$$

Для коэффициентов разложения функции
$$
\dfrac{1}{X(z)}=z-V_1+ \dfrac{X_1^*}{z}+\dfrac{X_2^*}{z^2}+
\dfrac{X_3^*}{z^3}+\cdots, \quad z \to \infty, \eqno{(6.17)}
$$
точно так же получаем:
$$
\exp(-V(z))=1-\dfrac{V_1}{z}+\dfrac{X_1^*}{z^2}+
\dfrac{X_2^*}{z^3}+\dots, \quad z \to \infty.
$$

Отсюда имеем:
$$
\dfrac{V_1}{z}+\dfrac{V_2}{z^2}+\dfrac{V_3}{z^3}+\dots=-
\ln\left(1-\dfrac{V_1}{z}+\dfrac{X_1^*}{z^2}+
\dfrac{X_2^*}{z^3}+\dots\right).
$$

После дифференцирования отсюда получаем:
$$
\left(\dfrac{V_1}{z^2}+\dfrac{2V_2}{z^3}+\dfrac{3V_3}{z^4}
+\dots\right)\left(1-\dfrac{V_1}{z}+\dfrac{X_1^*}{z^2}+
\dfrac{X_2^*}{z^3}+\dots\right)=
$$
$$
=\dfrac{V_1}{z^2}-\dfrac{2X_1^*}{z^3}-\dfrac{3X_3^*}{z^4}-
\dfrac{4X_3^*}{z^5}-\dots .
$$

Приравнивая коэффициенты при одинаковых степенях слева и справа,
получаем, что коэффициенты разложения (6.17) вычисляются
согласно равенствам
$$
\begin{array}{c}
X_1^*=-V_2+\dfrac{1}{2}V_1^2=-0.045, \\
X_2^*=-V_3+V_1V_2-\dfrac{1}{6}V_1^3=-0.023,\\
X_3^*=-V_4+V_3V_1+\dfrac{1}{2}V_2^2-\dfrac{1}{2}V_2V_1^2+
\dfrac{1}{24}V_1^4=-0.015,\cdots\\
%\cdots\cdots\cdots\cdots\cdots\cdots\cdots\cdots\cdots\cdots\cdots\\
%X_n^*=-V_{n+1}+\dfrac{n}{n+1}V_nV_1-\cdots-
%\dfrac{2}{n+1}V_2X_{n-2}-
%\dfrac{1}{n+1}V_1X_{n-1},\\
%\cdots\cdots\cdots\cdots\cdots\cdots\cdots
%\cdots\cdots\cdots\cdots\cdots\cdots
%\cdots\cdots\cdots\cdots\cdots\cdots\cdots\cdots
\end{array}
$$

Используя интегральные представления (6.10) и (6.13) для
вычисления коэффициентов разложений (6.16) и (6.17) можно
использовать следующие формулы:
$$
X_n=-\dfrac{1}{\pi}\int\limits_{0}^{1}X(\tau)\sin \zeta(\tau)
\tau^{n-1}\,d\tau,\quad n=1,2,3,\dots,
$$
причем $X_1=1$, и
$$
X_n^*=\dfrac{1}{\pi}\int\limits_{0}^{1}
\dfrac{\sin\zeta(\tau)}{X(\tau)}\tau^{n-1}\,d\tau, \quad
n=1,2,3,\dots.
$$

\begin{center}
\item{}\section{Разложение решения по собственным
функциям характеристического уравнения}
\end{center}

\markboth{}{РАЗЛОЖЕНИЕ ПО СОБСТВЕННЫМ ФУНКЦИЯМ}

Докажем, что наша граничная задача имеет единственное решение,
представимое в виде разложения по собственным функциям
характеристического уравнения:
$$
h(x,\mu)=h_{as}(x,\mu)+\int\limits_{0}^{1}\exp(-\dfrac{x}{\eta})
\varphi(\eta,\mu)a(\eta)d\eta,
\eqno{(7.1)}
$$
или:
$$
h(x,\mu)=h_{as}(x,\mu)+\dfrac{3}{4}
\int\limits_{0}^{1}\exp(-\dfrac{x}{\eta})\dfrac{\eta
a(\eta)d\eta}{\eta-\mu}+
$$
$$
+\exp(-\dfrac{x}{\mu})\dfrac{\lambda(\mu)a(\mu)}{1-\mu^2}H_{+}(\mu).
\eqno{(7.2)}
$$
В разложении (7.2) функция $\Theta_{+}(\mu)$ есть функция Хэвисайда,
$$
\Theta_+(\mu)=\left\{
\begin{array}{l}
1, \qquad \mu>0, \\0, \qquad \mu<0.
\end{array}
\right.
$$

Неизвестным в разложении является функция $a(\eta)$, называемая
коэффициентом непрерывного спектра. Разложение автоматически
удовлетворяет граничным условиям. Подставим $x=0$ в (7.2) и
левую часть заменим согласно граничному условию (5.8) приходим к
сингулярному интегральному уравнению с ядром Коши.
$$
2U_{sl}-2G_v\mu+\dfrac{3}{4}\int\limits_{0}^{1}\dfrac{\eta
a(\eta)d\eta}{\eta-\mu}+\dfrac{\lambda(\mu)a(\mu)}{1-\mu^2}=0
\qquad 0<\mu<1.
$$

Введём вспомогательную функцию
$$
N(z)=\dfrac{3}{4}\int\limits_{0}^{1}\dfrac{\eta
a(\eta)}{\eta-z}d\eta.
\eqno{(7.3)}
$$
аналитичную в комплексной плоскости с разрезом вдоль отрезка
[0,1] действительной оси. Ее граничные значения сверху $N^+(\mu)$
и снизу $N^-(\mu)$ в
интервале (0,1) связаны формулами Сохоцкого
$$
N^+(\mu)-N^-(\mu)=i\dfrac{3}{2}\pi\mu a(\mu), \quad
\dfrac{1}{2}[N^+(\mu)-N^-(\mu)]=N(\mu),
\eqno{(7.4)}
$$
где
$$
N(\mu)=\dfrac{3}{4}\int\limits_{0}^{1}\dfrac{\eta
a(\eta)}{\eta-\mu}d\eta, \qquad \mu \in (0,1),
$$
$N(\mu)$ - сингулярный интергал.

С помощью граничных значений функций $N(z)$ и $\lambda(z)$
сведем уравнение (7.4) к неоднородной краевой задаче Римана:
$$
\lambda^+(\mu)[N^+(\mu)+2U_{sl}-2G_v\mu]=
\lambda^-(\mu)[N^-(\mu)+2U_{sl}-2G_v\mu],\;
0<\mu<1.
\eqno{(7.5)}
$$

Рассмотрим соответствующую однородную краевую задачу Римана:
$$
\dfrac{X^+(\mu)}{X^-(\mu)}=\dfrac{\lambda^+(\mu)}{\lambda^-(\mu)},
\qquad 0<\mu<1.
\eqno{(7.6)}
$$

В качестве решения задачи (7.6) возьмем ограниченное в концевых
точках промежутка интегрирования (разреза) решение
$$
X(z)=\dfrac{1}{z}\exp(V(z)), \qquad
V(z)=\dfrac{1}{z}\int\limits_{0}^{1}\dfrac{\zeta(\tau)d\tau}{\tau-z},
$$
где
$$
\zeta(\tau)=\theta(\tau)-\pi, \qquad
\theta(\tau)=\arg \lambda^+(z),
$$
причём
$$
\theta(0)=0, \qquad \zeta(\tau)=\dfrac \pi 2 -
\arctg\dfrac{4\lambda(\tau)}{3\pi \tau(1-\tau^2)}.
$$

Неоднородную краевую задачу (7.5) с помощью однородной (7.6)
преобразуем к задаче определения аналитической функции по ее
нулевому скачку в интервале (0,1):
$$
X^+(\mu)[N^+(\mu)+2U_{sl}-2G_v\mu]=X^-(\mu)[N^-(\mu)+2U_{sl}-2G_v\mu].
\eqno{(7.7)}
$$

Учитывая поведение всех функций, входящих в краевое условие
(7.7), получим его общее решение:
$$
N(z)=-2U_{sl}+2G_vz+\dfrac{c_0}{X(z)},
\eqno{(7.8)}
$$
где $c_0$ - произвольная постоянная. Разложим эту функцию в ряд
$$
N(z)=-2U_{sl}+2G_v z+c_0 z(1-\dfrac{V_1}{z}+...),
\eqno{(7.9)}
$$
или, приводя подобные члены,
$$
N(z)=\Big(-2U_{sl}-V_1 c_0\Big) +z\Big(2G_v+c_0\Big)...,
\eqno{(7.10)}
$$
 Решение (7.8) имеет простой полюс в точке $z=\infty$, в то
время как функция $N(z)$, определенная равенством (7.3),
исчезает в бесконечности как $1/z$. Поэтому, чтобы функцию
$N(z)$, определенную равенством (7.8) можно было принять в
качестве функции $N(z)$, определенной равенством (7.3), устраним
у решения (7.8) полюс в точке $z=\infty$ равенством
$\lim\limits_{z\to\infty}\dfrac{N(z)}{z}=0$. Потребуем, чтобы
функция исчезала в бесконечности, т.е. $N(\infty)=0$.

На этом пути получаем, что $c_0=-2G_v$.

После подстановки в (7.10) получаем следующее выражение для
безразмерной массовой скорости газа:
$$
U_{sl}=V_1 G_v.
\eqno{(7.11)}
$$

Неизвестный коэффициент непрерывного спектра найдём, если
подставим решение (7.8) в формулу Сохоцкого (7.4). На этом пути
получаем:
$$
 a(\eta)=-\dfrac{4G_v}{3\pi \eta
i}\Big[\dfrac{1}{X^+(\eta)}-\dfrac{1}{X^-(\eta)}\Big]=\dfrac{8G_v
\sin \zeta (\eta)}{3 \pi \eta X(\eta)}. \eqno{(7.12)}
$$

Вычислим коэффициент вязкости $\eta$ квантового ферми--газа.
Исходя из определения вязкости, имеем:
$$
  \eta= - \dfrac{P_{xy}}{\left(\dfrac{dU_y}{dx}\right)_{\infty}} =
- \dfrac{m \displaystyle\int{fV_xV_yd\Omega}}{G_v},
  \eqno{(7.13)}
$$
где $g_v$ - градиент размерной скорости, определяемый следующим
соотношением $g_v = \nu_0 G_v$.

В результате получаем следующее выражение:

$$
\eta=\dfrac{2\pi(2s+1)m^4 8 l_1(\alpha)}{2\pi \hbar
v_0(\sqrt{\beta})^5 15}, \eqno{(7.14)}
$$
где
$$
l_1(\alpha)=\int\limits_{0}^{\infty}C \ln(1+\exp(\alpha-C^2))dC.
$$
Исходя из определений числовой плотности $N$,
$
N=\int f d\Omega,
$
нетрудно найти, что
$$
N=\dfrac{2\pi(2s+1)m^3l_0(\alpha)}{(2\pi \hbar)^3
(\sqrt\beta)^3},
$$
где
$$l_0(\alpha)= \int\limits_0^\infty \ln(1+\exp(\alpha-C^2))dC.
\eqno{(7.15)}
 $$

Подставляя (7.15) в (7.14), получаем коэффициент вязкости:
$$
\eta=\dfrac{8 \rho l_1(\alpha) }{15\nu_0 \beta l_0(\alpha)},
\eqno{(7.16)}
$$
где $ \rho=Nm$ .

\begin{center}
\item{}\section{Профиль массовой скорости и функция распределения в
полупространстве}
\end{center}

\markboth{}{МАССОВАЯ СКОРОСТЬ И ФУНКЦИЯ РАСПРЕДЕЛЕНИЯ}

Профиль безразмерной массовой скорости строится согласно (8.9)
$$
U_y(x_1)=U_y^*(x_1)=\dfrac{3}{8}\int\limits_{-1}^{1}
(1-\mu^2)h(x,\mu)\,d\mu.
$$
Согласно (8.5) получаем следующее распределение массовой скорости
классического газа в полупространстве:
$$
U_y^*(x)=U_{sl}+G_vx+\dfrac{3}{8}\int\limits_{0}^{1}
\exp({-\dfrac{x}{\eta}})\,a(\eta)\,d\eta,
\eqno{(8.1)}
$$
При выводе было учтено, что
$$
\int\limits_{-1}^{1} (1-{\mu'}^2)\varphi(\eta,\mu')\,d\mu'\equiv
1.
$$
В распределении (8.1) функция $a(\eta)$ определяется равенством
(7.12). В результате для профиля массовой скорости классического газа
получаем:
$$
\dfrac{U_y^*(x)}{G_v}=V_1+x-\dfrac{1}{2\pi i} \int\limits_{0}^{1}
\exp({-\dfrac{x}{\eta}})\left[\dfrac{1}{X^+(\eta)}-
\dfrac{1}{X^-(\eta)}\right]\dfrac{d\eta}{\eta},
\eqno{(8.2)}
$$
или
$$
\dfrac{U_y^*(x)}{G_v}=V_1+x+\dfrac{1}{\pi}\int\limits_{0}^{1}
\exp(-\dfrac{x}{\eta})\dfrac{\sin\zeta(\eta)}{\eta
X(\eta)}\,d\eta. \eqno{(8.3)}
$$

Интегралы из правых частей (8.2) и (8.3) при $x=0$ можно
вычислить аналитически. Для этого воспользуемся интегральным
представлением (6.11)
$$
\dfrac{1}{X(z)}-z+V_1=-\dfrac{1}{\pi }\int\limits_{0}^{1}
\left[\dfrac{\sin\zeta(\eta)}{ X(\eta)}\right]\dfrac{d\eta}
{\eta-z}.
\eqno{(8.4)}
$$
Устремляя в (8.4) $z \to 0$ вдоль отрицательной части
действительной оси, получаем, что
$$
-\dfrac{1}{\pi}\int\limits_{0}^{1} \left[\dfrac{\sin\zeta(\eta)}
{X(\eta)}\right]\dfrac{d\eta} {\eta}=\dfrac{1}{X(-0)}+V_1.
\eqno{(8.5)}
$$
Массовую скорость классического газа изотермического
скольжения у стенки (при
$x=0$) найдём, если подставим (8.5) в (8.2) при $x=0$
$$
U_y^*(0)=-\dfrac{1}{X(-0)}G_v.
$$

Как указывалось выше, из формулы для факторизации
дисперсионной функции вытекает, что $X(-0)=-\sqrt{5}$.

Следовательно, массовая скорость газа классического газа у стенки равна
$$
U_y^*(0)=\dfrac{1}{\sqrt{5}}G_v=0.4472G_v.
\eqno{(8.6)}
$$

%Коэффициент непрерывного спектра разложения (5.23) функцию
%$a(\eta)$ представим с помощью равенства (5.31) в виде:

%$$
% a(\eta)=\dfrac{4 c_0 \sin \zeta(\eta)}{3\pi \eta
%X(\eta)}. \eqno{(5.38)}
%$$
%Равенство (5.38) означает, что разложение (5.23) есть разложение
%в виде Лапласа "по синусам":

%$$
%h(x,\mu)=\dfrac{3}{4}\int\limits_{0}^{1}\exp(-\dfrac{x}
%{\eta})\varphi(\eta,\mu)\dfrac{4c_0 \sin \zeta (\eta)}{3 \pi
%\eta X(\eta)}d\eta,
%$$

%$$
%h(x,\mu)=\int\limits_{0}^{1}\exp(-\dfrac{x}
%{\eta})\varphi(\eta,\mu)\dfrac{c_0 \sin \zeta (\eta)}{ \pi \eta
%X(\eta)}d\eta,
%$$

Перейдем к рассмотрению функции распределения.

В случае изотермического скольжения для построения функции
$h(x,\mu)$ подставим коэффициент непрерывного спектра $a(\eta)$,
определяемый равенством (7.12), в разложение (7.2). Получаем
%$$
%h(\bm r,\bm
%C)=2C_y\Big[U_0+G_v(x-\mu)\Big]+C_y\psi(x,\mu).
%\eqno{(8.12)}
%$$
%$$\extrarowheight=11pt
%\begin{array}{c}\displaystyle
%-\dfrac{h(x,\mu)}{2G_v}=-V_1+\mu-x+\dfrac{1}{2 \pi i}
%\int\limits_{0}^{1}\exp({-\dfrac{x}{\eta}})
%\left[\dfrac{1}{X^+(\eta)}-
%\dfrac{1}{X^-(\eta)}\right]\dfrac{d\eta}{\eta-\mu}+
 % \\+\dfrac{2\lambda(\mu)}
 % {3\pi i\mu(1-\mu^2)}\left[\dfrac{1}
  %{X^+(\mu)}-\dfrac{1}{X^-(\mu)}\right]
 % \exp(-\dfrac{x}{\mu})H_+(\mu),
%\end{array}
%\eqno{(8.13)}
%$$
%или, после некоторых преобразований,
$$
\dfrac{h(x,\mu)}{2G_v}=V_1+x-\mu+
\dfrac{1}{\pi}\int\limits_{0}^{1}
\exp(-\dfrac{x}{\eta})\dfrac{\sin\zeta(\eta)}{\eta X(\eta)}
\dfrac{d\eta}{\eta-\mu}+$$$$+\dfrac{\cos\zeta(\mu)}{X(\mu)}
\exp(-\dfrac{x}{\mu})\Theta_+(\mu).
\eqno{(8.7)}
$$

Отсюда при $x=0$ имеем:
$$
\dfrac{h(0,\mu)}{2G_v}=V_1-\mu+\dfrac{1}{\pi}\int\limits_{0}^{1}
\dfrac{\sin \zeta(\eta)}{X(\eta)}\dfrac{d\eta}{\eta-\mu}+
\dfrac{\cos\zeta(\mu)}{X(\mu)}\Theta_+(\mu),\quad \mu\in (-1,1).
$$

Согласно интегральному представлению (6.11) первое слагаемое из
правой части последнего равенства равно
$$
\dfrac{1}{\pi}\int\limits_{0}^{1}
\dfrac{\sin\zeta(\eta)}{X(\eta)}\dfrac{d\eta}{\eta-\mu}=
-\dfrac{\cos\zeta_+(\mu)}{X(\mu)}+\mu-V_1.
$$

Следовательно, на границе полупространства функция распределения
вычисляется по формуле
$$
\dfrac{h(0,\mu)}{2G_v}=
\dfrac{\Theta_+(\mu)\cos\zeta(\mu)-\cos\zeta_+(\mu)}{X(\mu)},\quad
\mu\in (-1,1).
\eqno{(8.8)}
$$

Из формулы (8.8) видно, что при $0<\mu<1$\; $h(0,\mu)=0$,\;
 что
в точности совпадает с граничным условием (5.2).

При $-1<\mu<0$ из формулы (5.3) получаем функцию распределения
летящих к стенке молекул:
$$
h(0,\mu)=-\dfrac{2G_v}{X(\mu)}.
%\eqno{(8.15)}
$$

Отсюда видно, что
$$
h(0,\mu)= -\dfrac{2G_v}{X(\mu)}\Theta_+(-\mu),\quad \mu\in (-1,1).
$$

Из формулы (8.8) видно, что при $x\to +\infty$
$$
h(x,\mu)=2G_v(V_1+x-\mu)+o(1)=2U_{sl}+2G_v(x-\mu),
$$
что в точности совпадает с граничным условием (5.3) вдали от
стенки.

Формулу для скорости скольжения представим в размерном виде.
Учитывая, что
$ U_{sl}=\sqrt {\beta}u_{sl}$, получаем:
$
U_{sl}=V_1G_v=\sqrt{\beta}u_{sl}.
$
Следовательно,
$
u_{sl}=\dfrac{V_1}{\sqrt{\beta}}G_v,
$
где
$
g_v=\nu_0 G_v.
$
Отсюда имеем:
$$
u_{sl}=\dfrac{V_1}{\sqrt{\beta}l}G_vl= \dfrac{V_1}{\sqrt{\beta}l
\nu_0}g_vl.
$$
Учитывая то, что мы выбираем длину свободного пробега
согласно Черчиньяни \cite{94},
как $l(\alpha)=\dfrac{\eta}{\rho}\sqrt{\pi\beta} $,
получаем следующее:
$$
l \sqrt{\beta}=\dfrac{\eta}{\rho}\sqrt{\pi \beta}\sqrt{\beta}=
\dfrac{8\rho l_1(\alpha)}{15\nu_0 \beta
l_0(\alpha)}\dfrac{1}{\rho}\sqrt{\pi}\beta=
\dfrac{8\sqrt{\pi}l_1(\alpha)}{15\nu_0 l_0(\alpha)}.
$$
Возвращаюсь к выражению для размерной скорости скольжения,
учитывая полученное, приходим к следующему выражению:
$$
u_{sl}=K_v(\alpha)l g_v, \eqno{(8.9)}
$$
где
$$
K_v(\alpha)=\dfrac{15V_1\nu_0l_0(\alpha)}{8 \sqrt{\pi}l_1(\alpha)\nu_0}.
$$

Аналогично скорости скольжения, получим профиль размерной
массовой скорости в полупространстве. Исходя из того, что
$$
{U_y(x)^*}=V_1G_v+G_vx + \dfrac{1}{\pi }G_v \int\limits_{0}^{1}
\exp({-\dfrac{x}{\eta}})\left[\dfrac{\sin\zeta(\eta)}{
X(\eta)}\right]\dfrac{d\eta}{\eta},
$$
или, иначе:
$
{U_y^*(x)}=H(x,\alpha)G_v,
$
где
$$
H(x,\alpha)=V_1+x+\dfrac{1}{\pi } \int\limits_{0}^{1}
\exp({-\dfrac{x}{\eta}})\left[\dfrac{\sin\zeta(\eta)}{
X(\eta)}\right]\dfrac{d\eta}{\eta}.
$$
Получаем выражение для размерной массовой скорости газа:
$$
u_y(x)=\dfrac{U_y(x)}{\sqrt{\beta}}=
\dfrac{H(x,\alpha)}{\sqrt{\beta}}G_v(\alpha)=
\dfrac{H(x,\alpha)}{\sqrt{\beta}l}G_v(\alpha)l.
$$
Выполняя преобразования, аналогичные тем, которые использовались
 для получения размерной скорости скольжения, приходим к следующему выражению
для размерной массовой скорости:
$$
u_y(x)=\dfrac{H(x,\alpha)l g_v 15 l_0(\alpha)}{8
\sqrt{\pi}l_1(\alpha)}=K_{v}^{*}(x, \alpha)l g_v,
\eqno{(8.10)}
$$
где
$$
K_{v}^{*}(x,\alpha)=\dfrac{15H(x,\alpha)l_0(\alpha)}{8\sqrt{\pi}
l_1(\alpha)}.
$$

При $ \alpha\to -\infty$ (когда квантовый ферми--газ
переходит в больцмановский), получаем известный (см. \cite{76})
результат:
$
K_v(-\infty)=15V_1/8.
$
Отсюда получаем следующее выражение для скорости скольжения:
$$
u_{sl}=K_v^0l(\alpha)g_v,
$$
где $K_v^0$ -- коэффициент изотермическорго скольжения,
равный
$$
K_v^0=\dfrac{15}{8}\cdot\dfrac{1}{\sqrt{5}}= 0.8385.
$$


\addcontentsline{toc}{section}{Список литературы}

\begin{thebibliography}{99}
\normalsize
\markboth{СПИСОК ЛИТЕРАТУРЫ}{СПИСОК ЛИТЕРАТУРЫ}

\bibitem{4}{\it Больцман Л.} Лекции по теории газов. //
М.: Гостехиздат. 1956.
\bibitem {4a} Больцман Л. Избранные труды. - М.: Наука, 1984. - 590 с.

\bibitem{5}{\it Ван Кампен.} Дисперсионное уравнение для
волн в плазме. //Сб. статей под ред. Бернашевского Г.А. и
Чернова З.С. 1961. М.: ИИЛ.360 с. (с. 57--70).

\bibitem{7}
{\it Владимиров В.С., Жаринов В.В.} Уравнения математической
физики. //М.:Физматлит. 2000. 399 с.

\bibitem{8}
{\it Гахов Ф.Д.} Краевые задачи.// М.:Наука. 1977. 640 с.

\bibitem{9}{\it Гахов Ф.Д., Черский Ю.И.} Уравнения типа
свертки.// М.: Наука. 1978. 296 c.

\bibitem{13}{\it Карлеман Т.} Математические вопросы теории
газов. //М.:ИЛ. 1960.

\bibitem{14}{\it Кейз К.М., Цвайфель П.Ф.} Линейная теория
переноса.// М.:Мир. 1972. 384 с.

\bibitem{14a} {\it Квашнин А.Ю., Латышев А.В., Юшканов А.А.}
Задача Крамерса в квантовых ферми -- газах с частотой
столкновений, пропорциональной модулю скорости молекул. //Труды
ин--та Системного анализа РАН "Динамика линейных и нелинейных
систем". 2006. Том 25 (2). С. 69 -- 73.

\bibitem{14b} {\it Квашнин А.Ю., Латышев А.В., Юшканов А.А.}
Задача Крамерса в квантовых бозе -- газах с частотой
столкновений, пропорциональной модулю скорости молекул// Сб. тр.
"Фундаментальные физико -- математические проблемы …". Изд--во
МГТУ "Станкин" 2008. Вып. 11. С. 74 -- 79.

\bibitem{14c}{\it Квашнин А.Ю., Латышев А.В., Юшканов А.А.} Задача
Крамерса для ферми -- газа с зеркально-диффузным граничным
условием. //Труды ин--та Системного анализа РАН "Динамика
неоднородных систем". 2008. Том 32(3). С. 101-- 105.

\bibitem{14d}{\it Квашнин А.Ю., Латышев А.В., Юшканов А.А.}
Изотермическое скольжение ферми-газа с зеркально-диффузным
отражением от границы // Известия высших учебных заведений.
Физика. 2009. Т. 52. № 12. С. 3--7.

\bibitem{14e}{\it Квашнин А.Ю.} Изотерическое скольжение
квантового бозе-газа с диффузным отражением от границы// Вестник
Московского государственного областного университа. Серия
"Физика-математика". 2009. №3. С. 14--25.

\bibitem {14f}{\it Квашнин А.Ю., Латышев А.В., Юшканов А.А.}
 Изотерическое
скольжение квантового бозе-газа с зеркально-диффузным отражением
от границы // Физика низких температур, 2010. Т. 36, N 4. C.
413-417.

\bibitem{14g} {\it Киттель Ч.} Квантовая теория твердых тел.//
М.:Наука,1967.-- 792 с.

\bibitem {17}{\it Костиков А.А, Латышев А.В., Юшканов А.А.}
Задача Крамерса с аккомодационными граничными условиями для
квантовых ферми – газов// Физика низких температур. 2008. № 34 С.
914-941.

\bibitem{30}{\it Латышев А.В.}Применение метода Кейза к решению
 линеаризованного кинетического БГК уравнения в задаче о тепловом
  скачке// ПММ. 1990. т.54. выпуск 4. - С. 581-586.


\bibitem{31} {\it Латышев А.В., Юшканов А.А.}
Аналитические аспекты решения модельных кинетических
 уравнений// Теор. и матем. физика. 1990. Т.85. №3 (декабрь).
 С. 428 -- 442.

\bibitem{33} {\it Латышев А.В., Юшканов А.А.}
 Аналитическое решение задач скольжения бинарного газа//
Теор. и матем. физика. 1991. Т. 86. №3 (март). С. 402 -- 419.

\bibitem{34} {\it Латышев А.В., Юшканов А.А.}
Теория и точные решения задач скольжения бинарного
газа вдоль плоской поверхности// Ж. выч. матем. и матем.
физ. 1991. Т.31. № 8. С. 1201--1210.

\bibitem{35} {\it Латышев А.В., Юшканов А.А.}
Уравнения свертки в задаче о диффузионном скольжении
бинарного газа с
аккомодацией// Поверхность.
Рентгеновские, синхротронные и нейтронные исследования.
1991. № 1. С. 31--37.


\bibitem{39} {\it Латышев А.В., Юшканов А.А.}
 Аналитическое решение граничных задач для нестационарных
модельных кинетических уравнений// Теор. и матем. физика. 1992.
Т. 92. № 1 (июль). С. 127--138.

\bibitem{41} {\it Латышев А.В., Юшканов А.А.}
Аналитическое решение одномерной задачи об умеренно
сильном испарении (конденсации) в полупространстве//
Ж. прикл. мех. и техн. физики. 1993. Т. 34. № 1. С. 102--106.

\bibitem{42} {\it Латышев А.В., Юшканов А.А.}
 Аналитическое решение задачи о сильном испарении
(конденсации)// Известия РАН. Сер. МЖГ. 1993. № 6. С.143--155.



\bibitem{49} {\it Латышев А.В., Юшканов А.А.}
 Тепловое и изотермическое скольжение в новом модельном
  кинетическом уравнении Лиу// Письма в журнал техн.
  физики. 1997. Т. 23. № 14. С. 13 -- 16.

\bibitem{50} {\it Латышев А.В., Юшканов А.А.}
 Аналитическое решение задачи о скольжении газа с
использованием модельного уравнения Больцмана с частотой,
пропорциональной скорости молекул// Поверхность.
Рентгеновские, синхротронные и нейтронные исследования.
1997. № 1. С. 92 -- 99.

\bibitem{51} {\it Латышев А.В., Юшканов А.А.}
Тепловое скольжение для газа с частотой столкновений,
пропорциональной скорости молекул//
Инженерно -- физический журнал. 1998. Т. 71. № 2. Март -- Апрель.
С. 353 -- 359.



\bibitem{57} {\it Латышев А.В., Юшканов А.А.}
 Слабое испарение (конденсация) с произвольным
коэффициентом испарения в газах с постоянной частотой столкновений
молекул// Инженерно -- физический ж. 2000, март--апрель.
Т. 73. №3. С. 542--549.

\bibitem{58} {\it Латышев А.В., Юшканов А.А.}
 Аналитическое решение задач скольжения с использованием
нового кинетического уравнения//
Письма в ЖТФ. 2000. Т. 26, вып. 23. С. 16--23.


\bibitem{60} {\it Латышев А.В., Юшканов А.А.}
Аккомодационные двухмоментные граничные условия в
задачах о тепловом и изотермическом скольжениях//
Инженерно -- физический журнал. 2001. Т. 74. № 3. С. 63--69.

\bibitem{61} {\it Латышев А.В., Юшканов А.А.}
Влияние свойств поверхности на скольжение газа с
переменной частотой столкновений молекул//
Поверхность. Рентгеновские, синхротронные и нейтронные исследования.
2001. № 7. С. 79--87.


\bibitem{63} {\it Латышев А.В., Юшканов А.А.}
 Граничные задачи для квантового ферми -- газа//
Теор. и матем. физика. 2001. Т. 129. № 3.
С. 491--502.

 \bibitem{64} {\it Латышев А.В., Юшканов А.А.}
 Граничные задачи для квантового бозе -- газа//
 Известия вузов. Сер. Физика. 2002. № 6. С. 51--56.



\bibitem{68} {\it Латышев А.В., Юшканов А.А.}
Моделирование кинетических процессов в квантовых
бозе -- газах и аналитическое решение граничных задач//
Матем. моделирование. 2003. №5. С. 80--94.


\bibitem{69} {\it Латышев А.В., Юшканов А.А.}
Кинетическое уравнение для квантовых ферми -- газов и
аналитическое решение граничных задач//
Теор. м матем. физика. Т. 134. № 2, февраль, 2003. С. 310--324.


\bibitem{70} {\it Латышев А.В., Юшканов А.А.}
Аналитическое решение задачи о скачке температуры в
металле// Ж. техн. физики. 2003. Т. 73. Вып. 7. С. 37--45.


\bibitem{72} {\it Латышев А.В., Юшканов А.А.}
  Моментные граничные условия в задачах скольжения
разреженного газа// Изв. РАН. Сер. МЖГ. 2004. № 2. С. 193--208.


 \bibitem{73} {\it Латышев А.В., Юшканов А.А.}
  Метод решения граничных задач для кинетических
уравнений// Ж. выч. матем. и матем. физики. 2004.
Т. 44. № 6. с. 1107--1118.


\bibitem{74} {\it Латышев А.В., Юшканов А.А.}
Аналитическое решение граничных задач кинетической
теории. Монография. М.: Изд--во МГОУ. 2004. 286 с.

\bibitem{75} {\it Латышев А.В., Юшканов А.А.}
Аналитическое решение задачи о скачке концентрации при
испарении бинарной газовой смеси// Письма в ЖТФ.
2004. Т. 30. Вып. 24. С. 12--19.

\bibitem{76} {\it Латышев А.В., Юшканов А.А.}
Кинетические уравнения типа Вильямса и их точные решения.
Монография. -- М.: Изд--во МГОУ. 2004. 271 с.

 \bibitem{78} {\it Латышев А.В., Юшканов А.А.}
 Задача Смолуховского для электронов в металле// Теор. и матем. физика.
2005, январь, Т. 142. № 1. С. 92--111.

\bibitem{79} {\it Латышев А.В., Юшканов А.А.}
 Метод сингулярных интегральных уравнений в граничных
задачах кинетической теории// Теор. и матем.
физика. 2005. Т. 143(4). № 3. 855--870. (437--454).


\bibitem{82} {\it Латышев А.В., Юшканов А.А.}
Влияние коэффициента испарения на параметры газа
вблизи поверхности// Инженерно -- физический журнал. 2007.
Т. 80. № 1. С. 121--126.

 \bibitem{83} {\it Латышев А.В., Юшканов А.А.} Задача
 Смолуховского для вырожденных Бозе -- газов//
Теор. и матем. физика. 2008. Т. 154. № 7. С. 1-- 14.

\bibitem{85}{\it Лифшиц Е.М., Питаевский Л.П.}
Физическая кинетика. //М.: Наука, 1979.

\bibitem{88} {\it Мусхелишвили Н.И.} Сингулярные
интегральные уравнения. //М.: Наука, 1968.

%т
%у
\bibitem{91}{\it Ферцигер Дж., Капер Г.} Математическая
теория процессов переноса в газах.// М.: Мир, 1976.
%х
%ц
\bibitem{92}{\it Чепмен С., Каулинг Т.} Математическая теория
неоднородных газов. //М.:ИЛ, 1960.


\bibitem{93}{\it Черчиньяни К.} Математические методы в
кинетической теории газов. //М. : Мир, 1973.

\bibitem{94} {\it Черчиньяни К.} О методах решения
уравнения Больцмана// Неравновесные явления: Уравнение
Больцмана. -- М. Мир. 1986. C. 132-204.



\bibitem{95}{\it Черчиньяни К.} Теория и приложения
уравнения Больцмана.// М.: Мир, 1978.


\bibitem{96}{\it Шахов Е.М.} Метод исследования
движений разреженного газа. //М.: Наука, 1974.


\bibitem{97}{\it Халатников И. М. } Введение в теорию
сверхтекучести.// М.: Наука, 1965. 160 с.

\bibitem{100}{\it Bardos C., Caflish R., Nikolaenko B.}
The Milne and Kramers problems for the
Boltzmann equation of a hard sphere gas// Comm. Pure Appl.
Math. 1986. V. 39. P. 323--352.


\bibitem{103}{\it Bhatnagar P.L., Gross E.M., Krook M.}
Model for collision processes in gases.
I. Small amplitude processes in charged and neutral
one component systems// Phys. Rev. 1954. V. 94. P. 511--525.


\bibitem{104}{\it Case K.M. } Elementary solutions of the
transport equations and their applications// Ann. Phys. V.9. \No 1.
1960. P. 1--23.

\bibitem{104a}{\it Cassell J.S. and Williams M.M.R.} An Exact Solution of the
Temperature Slip Problem in Rarefied Gases// Transport Theory and Statistical
Physics, 2(1), 81--90 (1972).

\bibitem{105}{\it Cercignani C.} Elementary solutions
of the linearized gas -- dynamics Boltzmann equation and
their applications to the slip -- flow problem// Ann.
Phys.(USA) 1962. V. 20. \No 2. P. 219--233.

\bibitem{106}{\it Cercignani C.} The method of elementary
solutions for kinetic models with velocity-dependent
collision frequency// Ann. Phys. 1966. V. 40. P. 469--481.

\bibitem{106a} {\it Cercignani C. Sernagiotto F.} The method of elementary
solutions for time-dependent problems in linearized kinetic
theory// Ann. Phys. 1964. V. 30. P.154-167.

\bibitem{107}
{\it Cercignani C.} The Kramers problem for a not
completely diffusing wall// J. Math. Phys. Appl. 1965. V.10.
P. 568--586.

\bibitem{109}{\it Cercignani C., Foresti P., Sernagiotto F.}
Dependence of the slip coefficient on the form of the
collision frequency// Part 2. Nuovo Cimento. 1968. V.
LV11. B. No.2. P. 297--306.

\bibitem{110}{\it Cercignani C., Lampis M.} Kinetic model for
gas--surface ineraction// Transport Theory and Statist. Physics.
1971. V.1. P. 101--109.

\bibitem{110a} {\it Chapman S.} On the kinetic theory of a gas; Part 2,
A composite monatomic gas,
 diffusion viscosity and thermal conduction// Phil. Trans. Roy. Sos.
 London, 1917. v.217. p. 118.

\bibitem{111}{\it Diallo S. O.} Condensate fraction and atomic
kinetic energy
of liquid ${^3}$He-${^4}$He mixtures // Archiv: cond-mat/0609529.


\bibitem{113} {\it Frisch H.} Analytic solution of the
velocity -- slip and diffusion -- slip problems
by a Cauchy integral method// Transport Theory and Statist. Physics.
1988. V. 11. \No 2. P.
615--633.

\bibitem {118} {\it Enskog D.} Kinetische Theorie der Vorgunge in massing
verdunnten Gasen. Diss. Uppsala, 1917.

\bibitem{118a} {\it Kvashnin A.Yu.,Latyshev A.V., Yushkanov A.A. }
Isothermal slip of a Fermi gas with specular-diffuse reflection
from the boundary// Russian Physics Journal. Springer New York.
Volume 52, Number 12 / Декабрь 2009 г., C. 1251-1257.

\bibitem{Kundt} {\it Kundt A., Warburg E.}
\"{U}ber Reibung und W\"{a}rmeleitung verd\"{u}nner
Gase// Annalen der Physik. 1875. V. 232 (10), 177--211.


\bibitem{120}{\it Latyshev A.V., Yushkanov A.A.}
Analytic solutions of boundary value problem for model kinetic
equatins// Math. Models of Non--Linear Excitations, Transfer,
Dynamics, and Control in Condensed Systems and Other Media.
Edited by L.A. Uvarova and A.V. Latyshev. Kluwer Academic. New
York -- Moscow. 2001. P. 17--24.

\bibitem{120a} {\it Latyshev A.V.,Yushkanov A.A.}
Boundary value problems for a model Boltzmann
equation with frequency proportional to the molecule velocity//
Fluid Dynamics.- 1996.-V.31.-№ 3. - p. 454-466.

\bibitem{121}{\it Levin K., Qijin Chen}. Finite Temperature Effects in
Ultracold Fermi Gases // Archiv: cond-mat/0610006.

\bibitem{123}{\it Loyalka S.K.} Slip in the thermal creep
flow// Phys. Fluids. 1971. V. 14. No. 1. P. 21--24.

\bibitem{124} {\it Loyalka S.K.} Approximative
method in the kinetic theory// Phys. Fluids. 1971. V. 14.
\No 11. P. 2291--2294.


\bibitem{125}{\it Loyalka S.K., Cipolla J.W., Jr}. Thermal
creep sleep with arbitrary accomodation at the
surface// Phys. Fluids. 1971. V. 14. \No 8. P. 1656--1661.

\bibitem{127} {\it Maxwell J.C.} On the dynamical theory of gases// Phil.
Trans. Roy. Soc. London, 1867.

\bibitem{128}{\it Maxwell J.C.} Maxwell J.C.
Illustrations of the dynamical theory of gases. I.
on the motion and collisions of perfectly elastic spheres; II.
On the process of diffusion of
two or more kinds of moving particles among one another; III.
On the collision of perfectly elastic bodies of any form// Phil. Mag., 1860.


\bibitem{129}{\it Modugno G}. Fermi --- Bose mixture with tunable
interactions // Archiv: cond-mat/0702277.

\bibitem{130} {\it Pao Y.--P.} Some boundary value problems
in the kinetic theory of gases// Phys. Fluids. V. 14. \No 11. 1971. P. 2285--2290.


\bibitem{132}{\it Siewert C.E.} Kramers' problem for a
variable collision frequency model// Eur. J. Appl. Math.
2000. V. 12. C. 179 -- 189.


\bibitem{136}{\it Siewert C.E., Sharipov F.} Model equations
in rarefied gas dynamics: Viscous--slip and thermal--slip
coefficiens// Phys. Fluids. 2002. V. 14. №12. P.
4123--4129.


\bibitem{141} {\it Uehling E.A., Uhlenbeck G.E.}
Transport Phenomena in
Einstein --- Bose and Fermi --- Dirac Gases// Physical Review. 1933,
April, Vol. 43. P. 552 -- 561.

\bibitem{144}{\it Williams M. M. R.} Boundary--value problems
in the kinetic theory of gases. Part 1. Slip flows// J. Fluid. Mech.
1969. V. 36. Pt.1. P. 145--159.
\end{thebibliography}
\end{document}